\documentclass[]{tandf2}

\usepackage{subfigure}

\usepackage[lined,ruled,norelsize]{algorithm2e}
\usepackage{color}

\newcommand{\commentout}[1]{}

\usepackage[numbers,sort&compress]{natbib}
\bibpunct[, ]{[}{]}{,}{n}{,}{,}
\renewcommand\bibfont{\fontsize{10}{12}\selectfont}
\makeatletter
\def\NAT@def@citea{\def\@citea{\NAT@separator}}
\makeatother

\theoremstyle{plain}

\theoremstyle{definition}

\theoremstyle{remark}

\renewcommand{\vec}[1]{\mbox{\boldmath$#1$}}
\newcommand{\tensor}[1]{\mbox{\boldmath{\ensuremath{#1}}}}

\usepackage{subfigure}
\usepackage{subfigmat}
\usepackage{array}
\newcolumntype{C}[1]{>{\centering\let\newline\\\arraybackslash\hspace{0pt}}m{#1}}

\begin{document}

\def\rtn{\par \noindent }
\def\pskip{\rtn }

\def\aa{ Acta Astronautica  }
\def\aam{ Adv. Appl. Mech.  }
\def\aas{ Atomization and Sprays  }

\def\aiaap{ AIAA Paper  }
\def\aj{ AIAA Journal  }
\def\ajou{ Aeronaut. J.  }
\def\annr{ Ann. Rev. Fluid Mech. }
\def\taj{ Astrophys. J. } 	
\def\tajss{ Astrophys. J., Suppl. Ser.}
\def\as{ Astron. and Astrophys. }
\def\autcf{Reprinted by permission of Elsevier Science from }
\def\autcff{\copyright ~the Combustion Institute }
\def\autjfm{Reprinted with permission by Cambridge University Press }
\def\anm{ Appl. Num. Math. }

\def\bbpc{ Ber. Bunsenges. Phys. Chem. }
\def\bitnm{ BIT Numer. Math	 }

\def\canm{ Comm. Appl. Num. Meth. }
\def\ces{ Chem. Eng. Sci. }
\def\cf{ Combust. Flame }
\def\cfl{ Comput. Fluids }
\def\cip{ Comput. Phys. }
\def\cvs{ Comput. Vis Sci. }
\def\cmame{ Comput. Methods Appl. Mech. Eng. }
\def\cpam{ Commun. Pure Appl. Math. }
\def\cpc{ Computer Phys. Communications }
\def\cras{ C. R. Acad. Sci. }
\def\cst{ Combust. Sci. Tech. }
\def\ctm{ Combust. Theory and Modelling }
\def\ctrsp{ Proc. of the Summer Program }
\def\ctrarb{ Annual Research Briefs }
\def\ptrsla{Phil. Trans. R. Soc. London A }
\def\zmp{Z. Math. Phys. }
\def\jsiam{J. Soc. Indust. Appl. Math. }
\def\pla{Physics Letters A }
\def\ptrsa{Phil. Trans. R. Soc. A }

\def\csat{ Composites Science and Technology }              	
\def\csat{ Composites Sci. and Tech. }              			
\def\camcos{ Comm. Appl. Math Comput. Sci. }

\def\DHCRS{ Dynamics of Heterogeneous Combustion and Reacting Systems }

\def\ent{ Entropie }
\def\etme{ Eng. Turb. Modelling and Exp. } 
\def\exf{ Exp. Fluids }
\def\esaimmmn{ ESAIM Math. Model. Num. }

\def\fd{ J. Fluid Dynamics }
\def\ftac{ Flow, Turb. and Combustion }

\def\icmf{ Int. Conf. Multiphase Flow }
\def\ija{ Int. J. Aeroacoustics }
\def\ijav{ Int. J. Acoust. Vib. }
\def\ijcfd{ Int. J. Comput. Fluid Dynamics }
\def\ijcm{ Int. J. Comput. Methods }
\def\ijhff{ Int. J. Heat Fluid Flow }
\def\ijmf{ Int. J. Multiphase Flow }
\def\ijmp{ Int. J. Modern Physics C }
\def\ijscd{ Int. J. Spray Combust. Dynamics }
\def\ijnme{ Int. J. Numer. Meth. Eng. }
\def\ijnmf{ Int. J.~Numer. Meth. Fluids }
\def\ijhmt{ Int. J.~Heat and Mass Transfer }
\def\ijts{ Int. J. of Therm. Sci. }
\def\iecf{Ind. Eng. Chem., Fundam.}
\def\ijck{ Int. J. Chem. Kinet. }                                      		
\def\ijck{ International Journal of Chemical Kinetics }         	


\def\ja{ J. Aircraft }
\def\jacic{ J. Aeropace Comput. Inform. Comm. }
\def\jaes{ J. Aeronaut. Sci. }
\def\jam{ SIAM J. Appl. Math. }
\def\jnm{ SIAM J. Num. Math. }
\def\jna{ SIAM J. Numer. Anal. }
\def\jame{ J. Appl. Mech. }
\def\jamp{ J. Appl. Phys. }
\def\jars{ J.~American Rocket Society }
\def\jas{ J. Atmos. Sci. }
\def\jasa{ J. Acous. Soc. Am. }
\def\jchp{ J. Chem. Phys. }
\def\jcht{ J. Chem. Thermodynamics }
\def\jcp{ J.~Comput. Phys. }
\def\je{ J. Energy }
\def\jep{ J. Eng. Gas Turb. and Power }
\def\jfe{ J. Fluids Eng. }
\def\jfm{ J.~Fluid Mech. }
\def\jfs{ J. Fluids Struct. }
\def\jht{ J. Heat Trans. }
\def\jie{ J. Inst. Energy }
\def\jmta{ J. M\'ec. Th\'eor. Appl. }
\def\jpp{ J.~Prop.~Power }
\def\jrnbs{ J. Res. Natl. Bur. Stand. }
\def\jsc{ J. Sci. Comput. }
\def\sjsc{ SIAM J. Sci. Comput. }
\def\jsr{ J. Spacecrafts and Rockets }
\def\jsv{ J.~Sound Vib. }
\def\jssc{ SIAM J. Sci. Stat. Comput. }
\def\jt{ J.~Turb. }
\def\jtht{ J. Thermophysics and Heat Trans. }
\def\jtu{ J. Turbomach. }
\def\jpca{J. Phys. Chem. A}
\def\jpdap{ Journal of Physics D: Applied Physics }          	
\def\jpdap{ J. Phys. D: Appl. Phys. }          				

\def\moc{ Math. Comp. }
\def\mwr{ Mon. Weather Rev. }
\def\mnras{ Mon. Notices Royal Astron. Soc .}

\def\ned{ Nuclear Eng. and Design }

\def\paa{ Prog. in Astronautics and Aeronautics}
\def\pas{ Prog. Aerospace Sci. }
\def\pcfd{ Prog. Comput. Fluid Dynamics }
\def\pci{ Proc. Combust. Inst. }
\def\pcp{ Prog. Comput. Phys. }
\def\pecs{ Prog. Energy Comb. Sci. }
\def\pf{ Phys. Fluids }
\def\pieee{ Proc. IEEE. }
\def\pime{ Proc. Instn. Mech. Engrs. }
\def\plms{ Proc. London Math. Soc }
\def\ppsc{ Particle and Particle Systems Characterization }
\def\prl{ Phys. Rev. Lett. }
\def\prsl{ Proc. R. Soc. Lond. }
\def\prsla{ Proc. R. Soc. Lond. A }
\def\pt{ Powder Technology }
\def\pep{ Propellants, Explosives, Pyrotechnics }          	

\def\qjmam{ Q. J. Mech. Appl. Math. }
\def\qjrms{ Q. J. R. Meteorol. Soc. }

\def\ra{ La Rech. A\'{e }rospatiale }
\def\rpa{ Rev. Phys. Appl. }

\def\tcfd{ Theoret. Comput. Fluid Dynamics }
\def\tcsme{ ASME Trans. }
\def\ti{ Technique de l'Ing\'enieur }
\def\tsfp{ Turb. Shear Flow Phenomena }

\def\WSSCI{ WSS/CI }

\def\Litem{\par\noindent }
\font\smc=cmcsc10

\title{A Fourth-Order Adaptive Mesh Refinement Algorithm for the Multicomponent, Reacting Compressible Navier-Stokes Equations}
\author{
Matthew Emmett\textsuperscript{a}, Emmanuel Motheau\textsuperscript{b} $^{\ast}$\thanks{$^\ast$Corresponding author. Email: emotheau@lbl.gov},  Weiqun Zhang\textsuperscript{b}, Michael Minion\textsuperscript{b} and John B. Bell\textsuperscript{b} \\  \vspace{6pt} 
\affil{\textsuperscript{a} Computer Modeling Group, Calgary, 3710 33 Street NW, Alberta, T2L 2M1 Canada; \\ \textsuperscript{b}Center for Computational Sciences and Engineering, Computational Research Division, Lawrence Berkeley National Laboratory, Berkeley, CA 94720-8139, USA.}
}

\maketitle

\begin{abstract}
In this paper we present a fourth-order in space and time block-structured adaptive mesh refinement algorithm for the compressible multicomponent reacting Navier-Stokes equations. The algorithm uses a finite volume approach that incorporates a fourth-order discretization of the convective terms.
The time stepping algorithm is based on a multi-level spectral deferred corrections method that enables explicit treatment of advection and diffusion coupled with an implicit treatment of reactions. The temporal scheme is embedded in a block-structured adaptive mesh refinement algorithm that includes subcycling in time with spectral deferred correction sweeps applied on levels. Here we present the details of the multi-level scheme paying particular attention to the treatment of coarse-fine boundaries required to maintain fourth-order accuracy in time. We then demonstrate the convergence properties of the algorithm on several test cases including both nonreacting and reacting flows.
Finally we present simulations of a vitiated
dimethyl ether jet in 2D  and a turbulent
hydrogen jet in 3D, both with detailed kinetics and transport.
\end{abstract}


\section{Introduction}
\label{sec:introduction}

In this paper we present a new fourth-order adaptive mesh refinement (AMR) algorithm
for the multicomponent reacting compressible Navier-Stokes equations. The new algorithm combines high-order discretizations  in both space and time with block-structured adaptive mesh refinement, making it an ideal approach for  direct numerical simulation of combustion on modern HPC architectures. The  algorithm is based on an adaptive finite volume spatial discretization coupled to a spectral deferred corrections (SDC) temporal integration strategy. The SDC approach facilitates explicit discretization of convection and diffusion with implicit discretization of reactions, enabling the overall algorithm to treat stiff chemistry with a time-step set by CFL considerations for convection and diffusion. The method incorporates a new approach to AMR time-stepping that increases coupling between the levels and improves overall efficiency.

Compressible reacting flow, like many systems governed by PDEs, exhibits a range of dynamic scales in both space and time with the finest scales existing in only a small fraction of the total area of interest in the simulation. In this case, the use of local adaptation of the computational grid can reduce the total number of spatial degrees of freedom necessary to resolve the solution compared to a uniform or static mesh. In this paper, we focus on block-structured adaptive mesh refinement (AMR) in the context of finite-volume spatial discretizations. The first block-structured AMR method for hyperbolic problems is introduced by Berger and Oliger~\cite{Berger:1984}. A conservative version of this methodology for gas
dynamics was developed by Berger and Colella~\cite{Berger:1989} and extended to three dimensions by Bell et al.~\cite{Bell:1994}.
Block-structured AMR has been applied in a wide range of fields
including astrophysics, combustion, magnetohydrodynamics, subsurface flow, and shock physics, 
and there are a number of public-domain software frameworks available for developing applications.
See Dubey et al. \cite{DUBEY2014} for a recent survey of AMR applications and software.

The original conservative AMR strategy with subcycling, used in many of the previous citations, proceeds as follows. Considering a grid with only two levels for simplicity,
the solution is first advanced for one time step on the coarse grid.
Next, the coarse solution is interpolated in space and (possibly)
time to supply boundary data for refined regions of the grid.
Then the solution on the fine regions is computed using the supplied boundary data.
Once fine grid solutions have been advanced to the same time as coarse grids, the solution on coarse grid cells corresponding to fine regions are overwritten by an average of the fine grid solution.
Additionally, the solution in coarse cells neighboring fine grids are corrected using the fine grid fluxes
where available.  This step, referred to as \emph{re-fluxing}, is necessary for global conservation. This two-level approach can be applied recursively to multiple levels with specified refinement ratios in space and time between levels.

Most of the previous work on AMR has focused on second-order finite volume discretizations of advection,
diffusion and other processes using operator splitting or other second-order temporal integration approaches.
In this paper, our goal is to develop a high-order AMR algorithm that combines the advantages of high-accuracy adaptive spatial discretization with higher-order temporal integration methods appropriate for AMR structures.
McCorquodale and Colella \cite{mccorquodale:2011} introduced a fourth-order AMR algorithm for gas dynamics that combines a high-order reconstruction method for spatial discretization with an explicit fourth-order Runge-Kutta approach for temporal integration. 
In the present study, the goal is to model reacting flow with detailed chemistry that is potentially stiff on hydrodynamic time scales.
Consequently, an implicit / explicit (IMEX) temporal integration strategy that couples an implicit
treatment of kinetics with explicit treatment of hydrodynamics can avoid both the severe time step restriction due to explicit treatment of
kinetics and the need to solve global nonlinear implicit equations coupling hydrodynamics and kinetics in a fully implicit treatment.

The temporal method pursued here is based on spectral deferred corrections 
(SDC), which is an iterative approach to temporal integration
based on formulating temporal integration in terms of a spectral collocation formula and solving
the resulting system using an efficient iterative algorithm.
SDC has been shown to provide a flexible platform for higher-order temporal 
integration with IMEX and/or multirate features (see e.g. \cite{Minion:2003,Bourlioux:2003,BouzarthMinion:2011}).
The SDC approach has 
previously been extended to include the possibility of computing some of  the correction iterations on coarsened versions of the problem. These {\it multi-level} SDC methods (MLSDC) \cite{Speck:2015} aim to reduce the overall computational cost of 
SDC by reducing the cost of some iterations and also are the basis of recently developed time-parallel methods \cite{Emmett:2012}.
Here we couple the structure of MLSDC with AMR to develop an efficient fourth-order temporal integration method
on block-structured adaptive grids.
%
The result is a time-stepping strategy, referred to as Adaptive Multi-Level SDC (AMLSDC), in which
we iterate over the entire adaptive grid hierarchy in space and time in a manner similar to full approximation scheme (FAS)
multigrid.
In this approach, corrections from fine grids are directly coupled to the coarse-grid solution.  This leads to
a more accurate treatment of coarse-fine boundaries than traditional AMR time-stepping 
algorithms in which coarse-grids are advanced independent of the fine grids and the solutions are then synchronized through
refluxing.

The remainder of this paper is organized as follows. In Section~\ref{sec:governing_equations}
we summarize the multicomponent reacting compressible Navier-Stokes equations. Next, the single grid algorithm that will form the basis of the AMLSDC approach is introduced in Section~\ref{sec:single_grid_algo}.  In Section~\ref{sec:AMR}
an overview of AMR is provided as well as the multi-level SDC approach, followed by a description of the AMLSDC strategy. In Section~\ref{sec:results}, the convergence
properties of the methodology are demonstrated, and the ability of the AMLSDC method to treat complex reacting flows is illustrated by simulation of a two-dimensional dimethyl ether jet flame and a three-dimensional turbulent hydrogen jet flame. 


\section{Governing equations}
\label{sec:governing_equations}

The multicomponent reacting compressible Navier-Stokes equations for $N_s$ species are given by

\begin{align}
\frac{\partial \rho}{\partial t} + \nabla \cdot (\rho
    \vec{u})= { } & 0, \label{eq:NS:mass} \\
\frac{\partial \rho \vec{u}}{\partial t} + \nabla \cdot (\rho
    \vec{u} \otimes \vec{u}) + \nabla p= { } & \nabla \cdot  \tensor{\tau}
  , \label{eqs:NS:momentum} \\
\frac{\partial \rho Y_s}{\partial t} + \nabla \cdot (\rho Y_s \vec{u})
= { } & - \nabla \cdot \mathcal{F}_s +  \dot{\omega}_s  \hspace{1cm} s=1,2,\ldots,N_s, \label{eqs:NS:species}  \\
\frac{\partial \rho E}{\partial t} + \nabla \cdot [(\rho E + p)
  \vec{u}] = { } & \nabla \cdot (\lambda \nabla T) + \nabla \cdot
  (\tensor{\tau} \cdot \vec{u}) - \nabla \cdot \sum_s \mathcal{F}_s h_s, \label{eqs:NS:energy}
\end{align}
where $\rho$ is the density, $\vec{u}$ is the velocity, $p$ is the
pressure, $E$ is the total energy (kinetic, internal and chemical), $T$ is the temperature and $\lambda$ is the thermal conductivity. The viscous stress tensor is given by
\begin{equation}
\tensor{\tau} = \eta (\nabla \vec{u} + (\nabla \vec{u})^T) + (\xi - \frac{2}{3} \eta ) (\nabla \cdot \vec{u}) \mathbf{I},
\end{equation}
where $\eta$ and $\xi$ are the shear and bulk viscosities. For
each of the chemical species~$s$, $Y_s$ is the mass fraction,
$\mathcal{F}_s$ is the species diffusion flux, and
$\dot{\omega}_s$ is the production rate.
The enthalpy term $h_s$ is given by:
\begin{equation}
h_s = \int_{T_0}^T c_{p,s}~{\rm d} T + \Delta h^0_{f,s},
\end{equation}
where $\Delta h^0_{f,s}$ is the enthalpy of formation of the species $s$ at $T_0 = 298.15$K and $c_{p,s}$ is the specific heat capacity at constant pressure.


The system is closed by an equation of state (EOS)
that specifies $p$ as a function of $\rho$, $T$ and $Y_s$.  An ideal gas mixture for the EOS is assumed:
\begin{equation}
 p = \rho T \frac{\mathfrak{R}}{\overline{W}},
\end{equation} 
where $\mathfrak{R}$ is the universal gas constant and $\overline{W}$ is the mean molecular weight defined as
\begin{equation}
\overline{W} = 1/\sum_{s=1}^{N_s} \frac{Y_s}{W_s},
\end{equation}
where $W_s$ is the molecular weight of species $s$.

Here the Soret and Dufour effects are ignored, and a mixture model for species diffusion is employed.  With these approximations, the species diffusion flux is given by
\begin{equation}
\bar{\mathcal{F}_s}
  = - \rho D_s \left( \nabla X_s + (X_s-Y_s) \frac{\nabla p}{p} \right),
\label{eq:fbar}
\end{equation}
where $X_s$ and $D_s$ are the mole fraction and the diffusion
coefficient of species $s$, respectively.
Transport coefficients are computed using the {\tt EGLIB} library \cite{Ern:1995}.
In order for species diffusion to be
consistent with mass conservation, the fluxes are modified by adding a
correction term as follows,
\begin{equation}
\mathcal{F}_s = \bar{\mathcal{F}}_s - Y_s \sum_j \bar{\mathcal{F}_j}.
\label{eq:vcorr}
\end{equation}
Since $\sum_s Y_s = 1$, Eq. (\ref{eq:vcorr}) implies that $\sum_s \mathcal{F}_s =0$.  Note
that the equation of continuity Eq.~(\ref{eq:NS:mass}) becomes redundant when species diffusion is properly defined.

\section{Single level algorithm}
\label{sec:single_grid_algo}

In this section, we describe the single level method that forms the basis of the MLSDC algorithm,
with an emphasis on the IMEX algorithm used to treat stiff kinetics.
The goal here is to construct an algorithm that is fourth-order in both space and time.
First, the spatial discretization procedures for convective and diffusive operators are described
in \S\ref{subsec:spatial_discretization_procedures}. Then,
the IMEX SDC method is presented in \S\ref{subsec:SDC_simple},
followed by a discussion of the details of how reactions are treated in 
\S\ref{sub:react}.

\subsection{Spatial discretization procedures}
\label{subsec:spatial_discretization_procedures}


The methodology presented here is based on finite volume discretizations in which the solution is represented by the average
of the conserved variables over a finite volume cell.
In a finite volume discretization, the point value at the center of a cell and the cell average agree to second-order
accuracy.   Consequently these two quantities can be equated when designing lower-order numerical methods; however, when constructing
a fourth-order discretization, the difference between the cell average and the point value at the center of the cell
needs to be accounted for in the discretization.  

The spatial discretization of the advection terms in the algorithm
uses the conservative finite-volume WENO reconstruction presented in \cite{Titarev:2004}.
In this approach, the solution is reconstructed at the cell interfaces with a fifth-order WENO reconstruction.
The reconstructed solution is then interpolated to Gauss quadrature nodes on the faces where the flux 
is computed using an HLLC Riemann solver.
(A midpoint rule for integrating fluxes is not sufficiently accurate for fourth-order convergence.)
We note that although the solution is reconstructed at cell interfaces with fifth-order WENO procedure,
the method is formally fourth-order accurate because we use a fourth-order quadrature rule to integrate the flux
over faces.
 
The diffusion terms are discretized using standard finite volume techniques.  First, the cell-averaged conserved variables are
used to compute fourth-order approximations to point values at cell centers using the procedure outlined in 
McCorquodale and Colella \cite{mccorquodale:2011}.
These point values of conserved quantities are then used to compute primitive variables, $\rho$, $U$, $p$, $T$ and $Y_s$.
Explicit formulae are then used to compute derivatives needed to compute the diffusive fluxes at Gauss points on the
cell-faces directly.  Similarly, diffusion coefficients
are computed at cell centers using point values and are then interpolated to Gauss points.
We refer the readers to \cite{Shu:1997} for the formulae of high-order
polynomial based reconstruction procedures.

\subsection{Spectral Deferred Correction method on a single level}
\label{subsec:SDC_simple}

We adopt a method of lines approach to integrate the ODEs obtained from the finite volume spatial discretization discussed
above.
Here,  the goal is to accurately track chemical mechanisms that are stiff relative to hydrodynamic time scales, making
a purely explicit discretization impractical.  Instead, we  use an IMEX approach that
treats advection and diffusion explicitly while treating reactions implicitly.
We denote by $F(U(t))$ the spatial discretization of the system
where $U$ is the vector of conservative variables. (Here we have suppressed the explicit dependence of $F$ on $t$;
however, it is straightforward to include explicit dependence on $t$.)
The function
$F(U(t))$ is then split into stiff $(F_R)$ and non-stiff $(F_{AD})$ parts, such that
\begin{equation}
F(U(t)) = F_{AD}(U(t)) + F_R(U(t)),
\end{equation}
where $A$, $D$ and $R$ denote advection, diffusion and reaction, respectively.

SDC methods are based on the integral form of the solution of a generic ODE
\begin{align}
\frac{\partial U(t)}{\partial t} &= F \bigl(U(t)\bigr), \hspace{0.5cm} t \in \left[t^n, t^{n+1} \right]; \\
U(t^n) &= U^n,
\end{align}
as the integral
\begin{equation}
  U(t) = U^n + \int_{t^n}^{t^{n+1}} F \bigl(U(\tau)\bigr)~ {\rm d} \tau.
  \label{eqn:picard}
\end{equation}
To discretize the integral, a single time-step $\left[t^n, t^{n+1}\right]$ is divided into a set of $M$ sub-intervals, with  $M+1$ temporal nodes given by
\begin{equation}
t^n = t^{n,0} < t^{n,1} < \ldots < t^{n,M} = t^{n+1}.
\end{equation}
Note that for notational simplicity, $t^m = t^{n,m}$. 
Here, the temporal nodes $m$ are chosen to be the appropriate Gauss-Lobatto quadrature nodes. Thus, the discrete reformulation of Eq.~(\ref{eqn:picard}) for the unknowns $U^m \approx U(t^m)$ is given by the collocation equation 
\begin{equation}
  U^m = U^n + \Delta t \sum_{j=0}^{M} q_{m,j} F\bigl(U^{j}, t^{j}\bigr), \hspace{0.5cm} m=1, \ldots, M, 
  \label{eqn:disc_coll}
\end{equation}
where the quadrature weights $q_{m,j}$ are given by
\begin{equation}
  q_{m,j} \equiv \frac{1}{\Delta t} \int_{t^{n}}^{t^m} l_{j}(\tau)~{\rm d}\tau,
    \hspace{0.5cm} m=1,\ldots, M;  j=0,\ldots,M
      \label{eqn:quad_weights}
\end{equation}
and $(l_{j})_{j=0,\ldots,M}$ are the Lagrange interpolating polynomials determined by the collocation nodes.

Eq.~(\ref{eqn:disc_coll}) can be recast in matrix form as 
\begin{equation}
\mathbf{U} = \mathbf{U}^n + \Delta t \mathbf{Q}\mathbf{F}
\label{eqn:SDC_picard_matrix_form_eq} 
\end{equation}
where 
\begin{align}
\mathbf{U} =& \left[U^1, \ldots, U^M \right]^T, \\
\mathbf{F} =& \left[F(U^0,t^0), \ldots, F(U^M,t^M) \right]^T, \\
\mathbf{Q} =&
    \begin{bmatrix}
    q_{1,0} & q_{1,1} & \ldots & q_{1,M} \\
    q_{2,0} & q_{2,1} & \ldots & q_{2,M} \\
    \vdots & \vdots & \ddots & \vdots \\
    q_{M,0} & q_{M,1} & \ldots & q_{M,M} 
    \end{bmatrix}
\end{align}
and $\mathbf{U}^n = \left[U^n, \ldots, U^n \right]^T$. Here, $\mathbf{Q}$ is referred to
as the \emph{integration matrix} and is of size $M \times (M+1)$.

Equation (\ref{eqn:SDC_picard_matrix_form_eq}) is an implicit equation for the unknowns $\mathbf{U}$ at all of the quadrature nodes
and is equivalent to a fully implicit Gauss Runge-Kutta method. Indeed, since the integration matrix $\mathbf{Q}$ is dense,
each entry of $\mathbf{U}$ depends on all other entries of $\mathbf{U}$ through the function values in the vector $\mathbf{F}$.
SDC methods can be viewed as an iterative scheme 
to solve Eq.~(\ref{eqn:SDC_picard_matrix_form_eq})
based on lower-order substepping over the quadrature nodes, hence avoiding the need to solve an implicit equation coupling the solution
at all nodes (as would result from a direct application of a Newton type method to Eq.~(\ref{eqn:SDC_picard_matrix_form_eq}).
Here, the iteration is based on an explicit update for advection and diffusion and an implicit update for reactions, which gives
the iterative update equation
%
\begin{multline}
U^{m+1,k+1} = U^{m,k+1} + \Delta t^m \left[F_{AD} \left(U^{m,k+1}\right) - F_{AD} \left(U^{m,k}\right) \right] \\ +
\Delta t^m \left[F_{R} \left(U^{m+1,k+1}\right) - F_{R} \left(U^{m+1,k}\right) \right] + \Delta t S^{m+1,k}
\label{eqn:SDC_sweep}
\end{multline}
for $m=0,\ldots,M-1$, where
\begin{equation}
S^{m+1,k} = \approx \frac{1}{\Delta t}   \int_{t^m}^{t^{m+1}} F \left(U^{k}\right)~{\rm d} t 
\end{equation}
is computed by integrating the Lagrange interpolating polynomial.  Specifically
\begin{align}
S^{1,k} =& \sum_{j=0}^M q_{1,j}   F\left(U^{j,k} \right) \\
S^{m+1,k} =& \sum_{j=0}^M \left(q_{m+1,j} - q_{m,j} \right) F\left(U^{j,k} \right), \;\; m>1 .
\end{align}
%
The process of solving Eq.~(\ref{eqn:SDC_sweep}) is referred as an \emph{SDC sweep}.
Given an approximation to the solution, the SDC sweep updates the solution at all of the collocation nodes $t^m$.
When the SDC iterations converge, the solution converges to the 
the spectral collocation method determined by the quadrature nodes and hence is of order $2M$ with $M+1$ Gauss-Lobatto nodes.
In standard SDC methods for ODEs, the formal accuracy of the approximation improves by one order
for each sweep so that $2M$ iterations are required to recover $2M^{th}$-order accuracy; however, for stiff equations, this
convergence can in practice be much slower.  Here, our goal is to
use information from coarser grids to accelerate convergence, reducing the number of iterations needed at the finest grid levels.
In order to monitor convergence, the SDC residual at
the last node is computed as follows:
\begin{equation}
\mathfrak{R}^{M,k} = U^n + \mathbf{q}\cdot\mathbf{F}^{k} - U^{M,k},
\label{eqn:convergence_criteria}
\end{equation}
where $\mathbf{q}$ is the last row of $\mathbf{Q}$. The SDC iterations are terminated when $|\mathfrak{R}^{M,k}| < \varepsilon_{\rm SDC}$, where $| \cdot |$ is the $\mathcal{L}^2$-norm, or when the number of SDC iterations has reached a maximum value $K$.

\subsection{Treatment of reactions}
\label{sub:react}

As discussed above for the spatial discretization, approximating the reactions using the cell averages is not
fourth order accurate.  
There are several choices for how to address this issue.  Here we  reconstruct
high-order point values (as done for diffusion), compute an update due to reactions, and then construct
the integral average over each cell from the point values.  At coarse-fine grid boundaries,
space-time interpolated values are required in one ghost cell  to perform the interpolation.

The implicit equation for the reaction terms in (\ref{eqn:SDC_sweep}) is solved with  Newton's method using an analytical form of
the Jacobian matrix.
In combustion applications, chemical kinetics can be extremely stiff so that a sufficiently accurate
initial guess for Newton's method may not be available on the first SDC sweep.
Hence, we provide an option to use the stiff ODE integrator DVODE to sub-cycle the evaluation of the reaction term
during the first SDC sweep $(k=1)$ in order to obtain a sufficiently accurate initial guess.
We use a tolerance of $10^{-14}$ in all implicit solves, and in practice, the number of Newton iterations needed
decreases as the SDC iterations converge due to the increasingly good initial guess provided by the previous SDC iteration.

\section{Adaptive Mesh Refinement}
\label{sec:AMR}

This section presents the extension of the SDC method to an adaptive mesh refinement framework. First, the basic principles of AMR are reviewed in \S\ref{subsec:grid_hierharchy},
followed by the description of the MLSDC approach in \S\ref{subsec:MLSDC}.
The extension of the MLSDC method to the AMR framework, referred to as AMLSDC,
is then presented in \S\ref{subsec:AMR_MLSDC}.
Here, some care is needed in the computation of the numerical fluxes at the boundaries of grids of different
levels of resolution. An algorithmic representation of the entire AMLSDC strategy is then presented in \S\ref{subsub:algo_AMLSDC}.

\subsection{Grid Hierarchy}
\label{subsec:grid_hierharchy}

\begin{figure}[htbp]
  \begin{center}
  \includegraphics[width=0.5\textwidth]{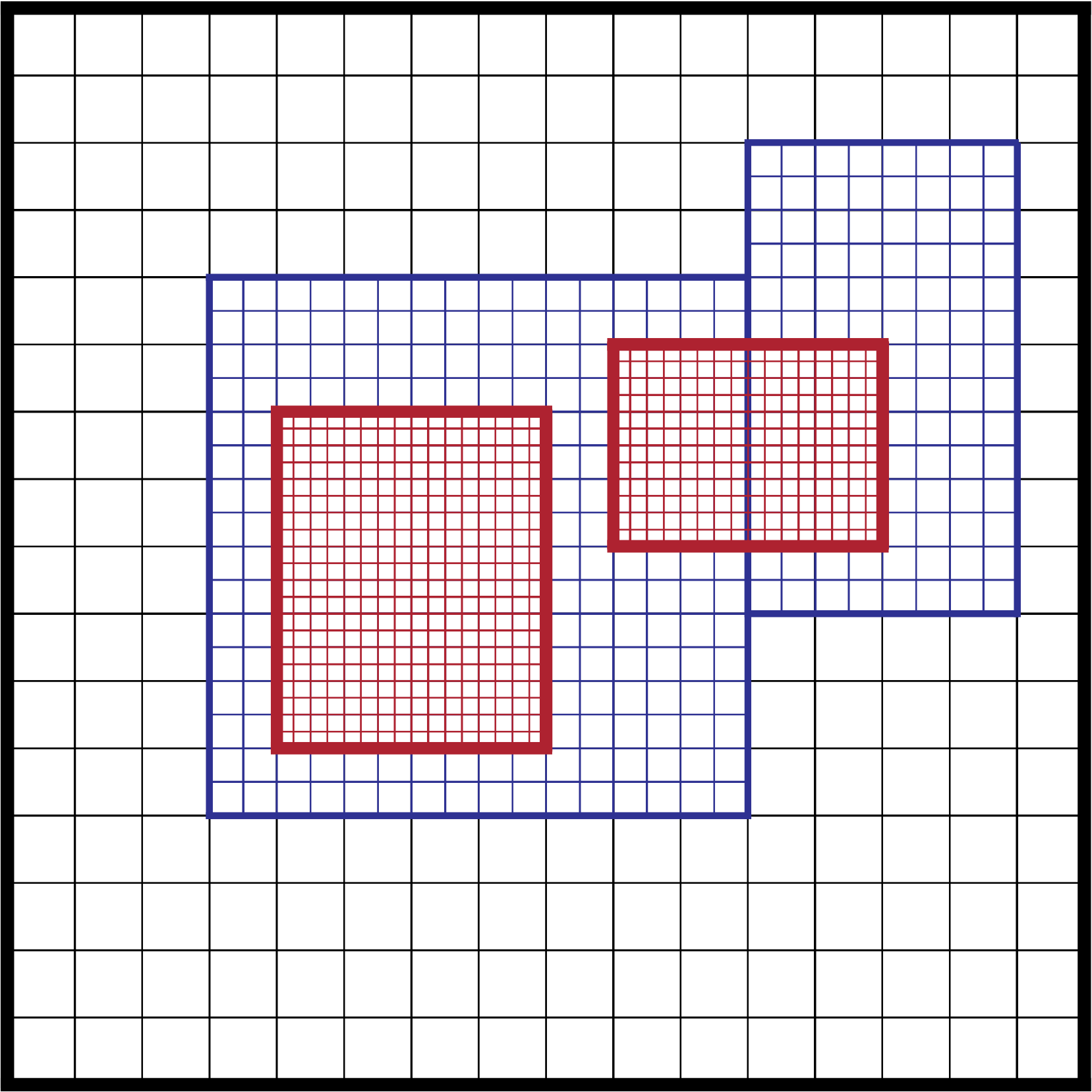}
  \end{center}
  \caption{Typical AMR grid structures in two dimensions.
  \label{fig:amr_grids}}
\end{figure}

As depicted in Figure~\ref{fig:amr_grids}, the computational domain
is represented
by a collection of grids at different levels of resolution.
The levels are denoted by $\ell=0, \ldots, L$. The entire computational domain is covered by the coarsest level ($\ell=0$); the finest level is denoted by $\ell = L$. The finer levels may or may not cover the entire domain; the grids at each level are properly nested in the sense
that the union of grids at level $\ell+1$ is contained within the union of grids at level $\ell$ buffered by a
layer of ghost cells except at physical boundaries.  Here the grid generation algorithm ensures that there
are at least four ghost cells between levels so that boundary data needed to advance level $\ell$ can be computed
from data at level $\ell-1$.
For ease of implementation of the interpolation procedures, the current algorithm assumes a ratio of $2$ in resolution between adjacent levels and that the cell size on each level is independent of direction. 

The grid hierarchy in AMR changes dynamically over time.
The initial grid hierarchy and subsequent regridding steps follow the procedure outlined in \cite{Bell:1994}: given grids at level $\ell$, an error estimation procedure is employed to tag cells where the error, as defined by user-specified routines, is above a given tolerance.
The specific criteria for refinement used here are discussed below.
The tagged cells are grouped into rectangular grids at level $\ell$ using the clustering algorithm given in \cite{Berger:1991}. These rectangular patches are refined to form the grids at level $\ell+1$. Large patches can be broken into smaller patches for distribution to multiple processors. The process is repeated until either an error tolerance criterion is satisfied or a specified maximum level is reached. In regions previously covered by fine grids the data are simply copied from old grids to new; in regions that are newly refined, data are interpolated from underlying coarser grids.
Note that in the present algorithm, the interpolation between grids is done with a
fourth-order stencil that conserves the sum of finite volumes. 
Furthermore, we enforce that regridding only occurs starting from level $\ell=0$; i.e., we only regrid
at the beginning of a coarse time step.

\subsection{Multi-Level Spectral Deferred Correction method}
\label{subsec:MLSDC}

Here we discuss the multi-level integration algorithm that forms the basis of the adaptive time-stepping algorithm.
In \cite{Emmett:2012,Speck:2015}, variations of SDC methods are
introduced in which SDC sweeps are performed on a hierarchy of different discretization levels.
The general strategy of multi-level SDC (MLSDC) methods is to reduce the computational cost per time step by replacing some of the SDC sweeps for a given problem by sweeps done on a coarsened version of the
problem. Solutions on different levels are coupled in the same
manner as in the Full Approximation Scheme (FAS) used in
multigrid methods for nonlinear problems \cite{Brandt:1977}. These FAS
corrections modify the coarse grid problem so that it converges to the
coarse representation of the fine grid solution.
We note that the MLSDC algorithm corresponds to the special case of the AMR algorithm in which the entire domain
is refined to the finest level.

\subsubsection{Node hierarchy and V-cycle algorithm}
\label{subsubsec:mlsdc_node_hierarchy}

\begin{figure}[t]
  \begin{center}
  \includegraphics[width=\textwidth]{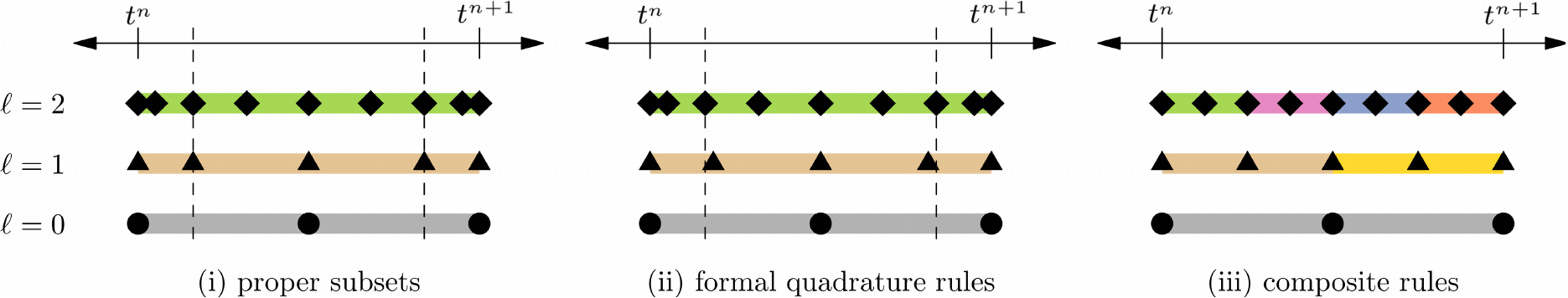}
  \end{center}
  \caption{Hierarchy of MLSDC collocation points for a three-level
    MLSDC method using 9 (diamonds), 5 (triangles), and 3 (circles)
    collocation points.  The coarsening strategies described in
    \S\ref{subsubsec:mlsdc_node_hierarchy} are shown: (i) proper
    subsets starting with 9 Gauss-Lobatto nodes, (ii) formal
    quadrature rules corresponding to 9, 5, and 3 point Gauss-Lobatto
    rules, and (iii) composite quadrature based on composing the 3
    point Gauss-Lobatto rule.  The composite quadrature rules are
    differentiated by the coloring of the nodes, which denote the span
     of the Lagrange polynomials used to compute quadrature weights.
    Note that, as shown, the only difference (highlighted by the
    dashed vertical lines) between the first two strategies occurs on
    level 2.}
  \label{fig:mlsdc_hierarchy}
\end{figure}

An MLSDC method is constructed by defining sets of collocation nodes, denoted by $\vec{t}_\ell$ for $\ell = 0, \ldots, L$, within a single time-step $\left[t^n, t^{n+1}\right]$. Each level $\ell$ is comprised of $M_\ell + 1$ collocation nodes so that $\vec{t}_\ell = \left[ t^0_\ell, \ldots, t_\ell^{M_\ell} \right]$ where $t^n =
t_\ell^0 < \cdots < t_\ell^{M_\ell} = t^{n+1}$. By convention, the first level $\ell=0$ is taken to be the coarsest (i.e., level $\ell=0$ has the fewest number of collocation nodes). There are several possible ways in which coarsening of the Gaussian integration nodes can be defined:

\begin{enumerate}
\item[(i)] they are proper subsets of fine nodes (i.e., $\vec{t}_\ell \subset \vec{t}_{\ell+1}$) but do not necessarily correspond to classical Gaussian quadrature rules; 
\item[(ii)] they correspond to Gaussian quadrature rules but are not necessarily proper subsets of fine nodes; 
\item[(iii)] they correspond to composite integration rules comprised of lower-order quadrature rules.
\end{enumerate}
These coarsening strategies are depicted in Figure~\ref{fig:mlsdc_hierarchy}.

In the present algorithm, a coarsening strategy based on composite integration rules is adopted since the formal order of quadrature on each level is the same, which facilitates restriction in time.  Furthermore, all examples presented in this paper use three Gauss-Lobatto quadrature nodes, or compositions thereof so that all quadrature rules are formally $4^{\rm th}$
order accurate.  For example, the composite node-to-node integration
matrices $\mathbf{Q}_0$ and $\mathbf{Q}_1$ corresponding to levels $0$ and $1$
depicted in Figure~\ref{fig:mlsdc_hierarchy}.(iii) are given by
\begin{equation}
  \mathbf{Q}_0 = \frac{1}{24}
  \begin{pmatrix}
    5 & 8 & -1 \\
    4 & 16 & 4
  \end{pmatrix}
\end{equation}
and
\begin{equation}
  \mathbf{Q}_1 =
   \frac{1}{48}
  \begin{pmatrix}
    5 & 8 & -1 & 0 & 0 \\
    4 & 16 & 4 & 0 & 0 \\
    4 & 16 & 9 & 8 & -1 \\
    4 & 16 & 8 & 16 & 4
  \end{pmatrix}
\end{equation}
respectively.

Unlike the single-level SDC method presented at \S\ref{subsec:SDC_simple}, an \emph{MLSDC iteration} between $k$ and $k+1$ proceeds by cycling through the levels using a multigrid V-cycle strategy \cite{Brandt:1977}. Basically, the idea is to ensure a two-way coupling between levels during an entire MLSDC iteration:
\begin{enumerate}
\item after updating the solution from $k$ to $k+1$ on a level $\ell$ with an SDC sweep, a correction term is computed and interpolated to fine level $\ell + 1$ in order to enhance the accuracy of the fine solution prior a new SDC sweep;
\item fine information is propagated to coarse levels through restriction and by applying a FAS correction term that will be detailed in \S\ref{subsubsec:FAS}. 
\end{enumerate}

When used in the context of AMR, data from coarse grids is required to provide boundary conditions for finer
levels.  
This necessitates that the iteration between levels start at the coarsest level, $\ell = 0$.
Note that this differs from the algorithm presented in \cite{Speck:2015}, where the authors start from the finest level first. Note also that as the present paper focuses on the extension of the method to AMR, the specific details regarding interpolation between levels as well as the computation of correction terms will be presented in \S\ref{subsubsec:mlsdcinterp}.

\subsubsection{Full Approximation Scheme (FAS) correction}
\label{subsubsec:FAS}

In MLSDC, the goal is to use coarser levels in both space and time to accelerate the convergence
of the SDC iteration on finer levels.
Because the system is nonlinear, a correction term is added to the coarse grid to represent the discrepancy between
the coarse and fine grid solutions so that when the iteration is converged, the coarse-grid residual vanishes.
The correction term for the Full Approximation Scheme (FAS) coarse MLSDC levels is derived in \cite{Emmett:2012,Speck:2015}.
Using the previous notation, the equation to be solved on the finest level $L$ is
\begin{equation}
\mathbf{U}_L = \mathbf{U}_L^n + \Delta t \mathbf{Q}_L\mathbf{F}_L.
\end{equation}
For FAS we assume that we have a number of coarser levels $\ell = 0,...,L-1$ in space and time and
corresponding discretizations.
Then given an approximate solution $\mathbf{U}_{\ell}$, the corresponding residual equation is 
\begin{equation}
\mathbf{U}_\ell + \delta \mathbf{U}_\ell - \Delta t \mathbf{Q}_\ell\mathbf{F}_\ell(\mathbf{U}_\ell+\delta \mathbf{U}_\ell)
=\mathbf{U}_\ell - \Delta t \mathbf{Q}_\ell\mathbf{F}_\ell(\mathbf{U}_\ell) + \mathbf{R}_\ell
\label{eqn:residual_fas}
\end{equation}
where $\delta \mathbf{U}_{\ell}$ is the correction and
$\mathbf{R}_{\ell} = \mathbf{U}_\ell^n + \Delta t \mathbf{Q}_\ell\mathbf{F}_\ell(\mathbf{U}_\ell) - \mathbf{U}_\ell$
is the residual. In nonlinear multigrid,
the residual equation is approximated at level $\ell$ by replacing the coarse residual $\mathbf{R}_{\ell}$
by the restriction $\mathcal{R}_{\ell+1}^{\ell} \mathbf{R}_{\ell+1}$ of the fine residual $\mathbf{R}_{\ell+1}$,
where $\mathcal{R}_{\ell+1}^{\ell} $ is the restriction operator between levels $\ell+1$ and $\ell$.
If we note that $\mathbf{U}_{\ell}^n = \mathcal{R}_{\ell+1}^{\ell}  \mathbf{U}_{\ell+1}^n$ and
$\mathbf{U}_{\ell} = \mathcal{R}_{\ell+1}^{\ell}  \mathbf{U}_{\ell+1}$, then
Equation~(\ref{eqn:residual_fas}) corresponds to a modified equation
\begin{equation}
\mathbf{U}_\ell + \delta \mathbf{U}_\ell - \Delta t \mathbf{Q}_\ell\mathbf{F}_\ell(\mathbf{U}_\ell+\delta \mathbf{U}_\ell)
= \mathbf{U}_\ell^n + \vec{\tau}_{\ell}
\end{equation}
where $\vec{\tau}_{\ell}$ is the FAS correction on the coarse level given by
\begin{equation}
\vec{\tau}_{\ell} = \Delta t \left( R_{\ell+1}^{\ell} \mathbf{Q}_{\ell+1} \mathbf{F}_{\ell+1}\left(  \mathbf{U}_{\ell+1} \right) - \mathbf{Q}_{\ell} \mathbf{F}_{\ell } \left( R_{\ell+1}^{\ell} \mathbf{U}_{\ell+1} \right)\right).
\end{equation}
Note that in the context of MLSDC the restriction operator $R_{\ell+1}^{\ell}$ incorporates both temporal (between MLSDC node hierarchies) and spatial restriction.  Specific details regarding time and space restriction in the AMR setting
will be presented in \S\ref{subsubsec:mlsdc_restrict_FAS}.

\subsection{Extension to Adaptive Multi-Level Spectral Deferred Correction method}
\label{subsec:AMR_MLSDC}

\begin{figure}[h]
  \begin{center}
  \includegraphics{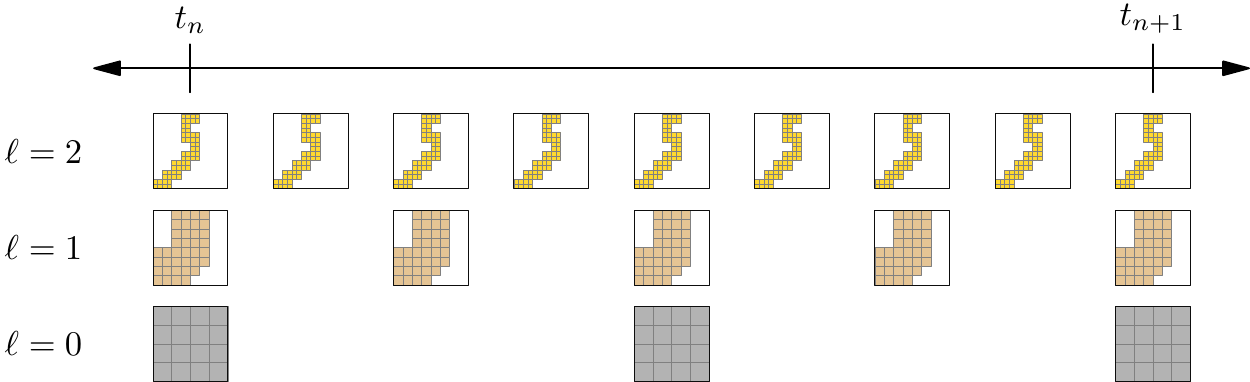}
  \end{center}
  \caption{AMLSDC hierarchy.  Each AMR level is associated with its own MLSDC level. The coarsest AMR level is integrated with 3 Gauss-Lobatto nodes. Subsequent AMR levels are integrated with increasing numbers of composite Gauss-Lobatto nodes.}
  \label{fig:mlsdcamr}
\end{figure}

The MLSDC strategy described in \S\ref{subsec:MLSDC} provides the basis for a new approach
to time integration in AMR algorithms.  The new algorithm, denoted
Adaptive Multi-Level Spectral Deferred Correction (AMLSDC),
couples AMR levels within an MLSDC time step iteration.
As depicted in Figure~\ref{fig:mlsdcamr}, the  coarsest AMR level becomes the coarsest MLSDC level,
and subsequent AMR levels are integrated with increasing numbers of quadrature nodes. The procedure followed is very similar to the MLSDC algorithm presented in \S\ref{subsubsec:mlsdc_node_hierarchy}: to evolve the system within one time step, several V-cycles through the AMLSDC hierarchy are performed, with FAS corrections updated as the iterations proceed. 

In the standard AMR time step approach, the solution is computed at level $\ell$ and then the fine solution at level
$\ell+1$ is advanced using boundary conditions interpolated in space and time from the coarser levels.
Here, we perform SDC sweeps on different levels in a manner similar to a multigrid V-cycle.
To preserve the fourth-order accuracy
of the single grid algorithm we require a fourth-order spatial interpolation and a temporal interpolation
that is consistent with the underlying quadrature rules. As noted earlier, we begin the V-cycle at the coarsest level
to provide ghost-cell data for finer levels.
The composite quadrature rules used for the temporal integration facilitate the construction of the restriction operator
and the FAS correction.

\subsubsection{Space and time interpolation for interior nodes and boundary ghost cells}
\label{subsubsec:mlsdcinterp}

Recall that in the MLSDC approach, a space and time interpolation is performed  on a correction term in order to improve the solution at finer levels before performing the subsequent SDC sweep. There are several ways that one can interpolate between levels. It has been found empirically that the method adopted and outlined below minimizes the number of AMLSDC V-cycles required to reach convergence on the test cases considered. As mentioned before, care must be taken concerning ghost-cells, and the interpolation procedure between levels $\ell$ and $\ell+1$ differs from the procedure used at interior nodes as explained below. 

\paragraph{Interior nodes}
\label{par:interior_nodes}

The fine level interior cells (i.e., non-ghost cells) are updated according to the  following procedure:
\begin{enumerate}
\item Compute the differences $\Delta \mathbf{U}_{\ell} \equiv
  \mathbf{U}_{\ell}^{k+1}  - \mathbf{U}_{\ell}^{k}$ at each coarse node in
  $\vec{t}_{\ell}$.
\item Interpolate $\Delta \mathbf{U}_{\ell}$ in space (here a high-order conservative quartic polynomial interpolation is employed) at each coarse node in $\vec{t}_{\ell}$ to obtain $\Delta \mathbf{U}_{\ell+1}$.
\item Interpolate $\Delta \mathbf{U}_{\ell+1}$ in time (using polynomial interpolation), from the coarse nodes in $\vec{t}_{\ell}$ to all fine
  nodes in $\vec{t}_{\ell+1}$ to obtain $\Delta \mathbf{U}_{\ell+1}^{*}$.  At this point $\Delta \mathbf{U}_{\ell+1}^{*}$ represents a coarse correction defined at all fine Gauss-Lobatto nodes.
\item Update fine interior cells according to $ \mathbf{U}_{\ell+1}^{k}
  \leftarrow \mathbf{U}_{\ell+1}^{k} + \Delta \mathbf{U}_{\ell+1}^{*}$ at each fine node in $\vec{t}_{\ell+1}$.
\end{enumerate}

At this point the interior cells of the fine solution at each fine temporal node have been updated with interpolated corrections from the grids at the next coarsest level. Again, this corresponds to conservative interpolation in space followed by polynomial interpolation in time. 
Specifically, interior cells on the fine level $\ell=1$ at quadrature points corresponding to $m=1,2,3,4$ are updated using
\begin{equation}
  \begin{aligned}
U_{\ell+1}^{1,k} & \leftarrow  {U}_{\ell+1}^{1,k}  + \frac{3}{8} \mathcal{I}_{\ell}^{\ell+1} \left({U}_{\ell}^{0,k+1} - {U}_{\ell}^{0,k}   \right), \\ 
& + \frac{6}{8} \mathcal{I}_{\ell}^{\ell+1} \left({U}_{\ell}^{1,k+1} - {U}_{\ell}^{1,k}   \right) - \frac{1}{8} \mathcal{I}_{\ell}^{\ell+1} \left({U}_{\ell}^{2,k+1} - {U}_{\ell=0}^{2,k}   \right), \\
{U}_{\ell+1}^{2,k} & \leftarrow  {U}_{\ell+1}^{2,k}  +  \mathcal{I}_{\ell}^{\ell+1} \left({U}_{\ell}^{1,k+1} - {U}_{\ell}^{1,k}   \right), \\
{U}_{\ell+1}^{3,k} & \leftarrow  {U}_{\ell+1}^{3,k}  - \frac{1}{8} \mathcal{I}_{\ell}^{\ell+1} \left({U}_{\ell}^{0,k+1} - {U}_{\ell}^{0,k}   \right), \\
& + \frac{6}{8} \mathcal{I}_{\ell}^{\ell+1} \left({U}_{\ell=0}^{1,k+1} - {U}_{\ell=0}^{1,k}   \right) + \frac{3}{8} \mathcal{I}_{\ell}^{\ell+1} \left({U}_{\ell=0}^{2,k+1} - {U}_{\ell=0}^{2,k}   \right), \\
{U}_{\ell+1}^{4,k} & \leftarrow  {U}_{\ell+1}^{4,k}  +  \mathcal{I}_{\ell}^{\ell+1} \left({U}_{\ell}^{2,k+1} - {U}_{\ell}^{2,k}   \right),
  \end{aligned}
  \label{eqn:exemple_interp_U}
\end{equation}
where $\mathcal{I}_{\ell}^{\ell+1}$ is the spatial interpolation operator between level $\ell$ to $\ell+1$.

\paragraph{Ghost-cells}
\label{par:ghost_cells}

Because an SDC sweep only advances the solution on interior cells, there is no solution in the ghost-cells at collocation nodes $\vec{t}_{\ell + 1}$ that can be corrected with the term $\Delta \mathbf{U}_{\ell}$. The strategy adopted here is that an equation similar to Eq.~(\ref{eqn:disc_coll}) can be solved in the ghost-cells, where the initial solution is interpolated in space from the  solution at iteration $k$ on level $\ell$, and the function values are interpolated in space and time but from values freshly updated at $k + 1$ on level $\ell$. Consequently, the values in ghost-cells at collocation nodes $\vec{t}_{\ell + 1}$ are expressed as follows:
\begin{equation}
  U^{m,k}_{\ell +1} = \mathcal{I}_{\ell}^{\ell+1} \left( U^{0,k}_{\ell}\right) + \Delta t \sum_{j=0}^{M_{\ell}} \bar{q}_{\ell,m,j} \mathcal{I}_{\ell}^{\ell+1} \left( F\bigl(U^{j,k+1}_{\ell}\bigr)\right), \hspace{0.5cm} m=1, \ldots, M_{\ell}, 
  \label{eqn:ghost_cells_eval_func}
\end{equation}
where $\bar{q}_{\ell,m,j}$ are quadrature weights for integrating the coarse-grid interpolating polynomial to the fine grid quadrature points. 
It should be noted that the values in the ghost-cells could be simply filled by the interpolated initial solution, viz. $U^{m,k}_{\ell +1} = \mathcal{I}_{\ell}^{\ell+1} \left( U^{m=0,k}_{\ell}\right)$
for all $m$.
However, it has been found that the use of
Eq.~(\ref{eqn:ghost_cells_eval_func}) improves
the convergence of AMLSDC iterations significantly.
This observation highlights the crucial role of boundary conditions at borders
of AMR patches and the need to impose values in ghost-cells consistent with the collocation node hierarchy.

\subsubsection{Restriction and FAS correction term}
\label{subsubsec:mlsdc_restrict_FAS}

As explained in \S\ref{subsubsec:FAS}, the FAS correction term $\tau$ is computed to
represent the difference between the solution at level $\ell$ and the finer solution at level $\ell+1$.
This procedure requires the application of a restriction operator. Recall that in the present algorithm,
the node hierarchy is built with a composite rule (see \S\ref{subsubsec:mlsdc_node_hierarchy}),
which facilitates the construction of the restriction operator in space and time.
Indeed, for each coarse collocation node in $\vec{t_{\ell}}$,
the coarse solution $U_{\ell}^{m}$ in a cell is set to be the average of the cells at level $U_{\ell+1}^{2m}$ that
cover the coarse cell. Consequently, the restriction operator is simply point-injection in time and averaging in space.

The restriction of the solution and the FAS correction term are then computed according to following procedure:
\begin{enumerate}
\item On level $\ell$, for each collocation node, for cells covered by fine cells at level $\ell+1$,
the solution $U_{\ell}^{m,k+1}$ is updated as follows:
\begin{equation}
U_{\ell}^{m,k+1} \leftarrow \mathcal{R}_{\ell+1}^{\ell} U_{\ell+1}^{2m,k+1}.
\label{eqn:restrict_U} 
\end{equation}
\item The coarse function evaluations on level $\ell$ are integrated over the quadrature points as follows:
\begin{equation}
I_{\ell}^{m,k+1} = \Delta t \sum_{j=0}^{M_{\ell}} q_{\ell,m,j}  F \left (U^{j,k+1}_{\ell} \right) \approx \int_{t^n}^{t^m} F\left(U^{m,k+1}_{\ell} \right)~{\rm d}t,
\label{eqn:start_FAS}
\end{equation}
where $q_{\ell,m,j}$ represents the entry of the quadrature matrix at level $\ell$.
\item Similarly, the fine function evaluations on level $\ell+1$ are integrated over the quadrature points as follows:
\begin{equation}
I_{\ell+1}^{m,k+1} =  \Delta t \sum_{j=0}^{M_{\ell+1}} q_{\ell+1,m,j}  F \left (U^{j,k+1}_{\ell+1} \right)   \approx \int_{t^n}^{t^m} F\left(U^{m,k+1}_{\ell+1} \right)~{\rm d}t.
\end{equation}
\item The fine integral $I_{\ell+1}^{m,k+1}$ is restricted in both time and space to level $\ell$ in order to obtain $I_{\ell^*}^{m,k+1}$, so that
\begin{equation}
I_{\ell^*}^{m,k+1}  = \mathcal{R}_{\ell+1}^{\ell} I_{\ell+1}^{2m,k+1}.
\end{equation}
\item Finally the FAS correction term $\tau_{\ell}^{m}$ is computed as follows:
\begin{equation}
\tau_{\ell}^{m} = I_{\ell^*}^{m,k+1}  - I_{\ell}^{m,k+1} .
\label{eqn:end_FAS}
\end{equation}
\end{enumerate}

\subsection{Adaptive Multi-Level Spectral Deferred Correction algorithm}
\label{subsub:algo_AMLSDC}

In this section, an algorithmic overview of the AMLSDC algorithm is provided.  We first describe an SDC sweep,  then
show how the SDC sweeps are organized into a V-cycle. Finally, we present the overall AMLSDC time step algorithm.
The SDC sweep procedure presented in \S\ref{subsec:SDC_simple} is shown here in Algorithm~\ref{alg:sdcsweep}.
Basically, within one AMLSDC iteration from $k$ to $k+1$ on a specific level $\ell$,
the SDC sweep updates the solution at each temporal collocation node $m$. 
Note that in the description of this algorithm, the superscript $k$ refers to the current approximation, which 
may include updates from coarser or finer grids.  


The procedure followed during an AMLSDC iteration from $k$ to $k+1$ is described in Algorithm~\ref{alg:amlsdc}.
It starts with the solution and function evaluations from the previous iteration $k$, and performs a V-cycle through the grid hierarchy, from  the coarsest level to the finest level, and then back from the finest to the coarsest level.

The Algorithm~\ref{alg:whole_amlsdc_RNS} presents the overall structure of
the procedure to advance from a time-step $t^n$ to $t^{n+1}$ on the coarsest level.
After each AMLSDC iteration over $k$, the convergence is checked with Eq.(\ref{eqn:convergence_criteria}).
If $|\mathfrak{R}^{(k+1)}| < \varepsilon_{\rm SDC}$ or if the max number $K$ of AMLSDC iterations is reached, viz. $k=K$, a last SDC sweep if performed over all the levels of the grid hierarchy
with $\bm{\tau}^{k}_{L}=0$
to enforce discrete conservation.

\begin{algorithm}[t]
  \SetKwComment{Comment}{\# }{}
  \SetCommentSty{textit}
  \DontPrintSemicolon

  \KwData{Solution $\mathbf{U}^{k}$ and function values  $F_{AD}\left(\mathbf{U}^{k}\right)$ and $F_R\left(\mathbf{U}^{k}\right)$ for each collocation node $t^m$, and (optionally) FAS corrections $\bm{\tau}^{k}$.}
  \KwResult{Solution $\mathbf{U}^{k+1}$ and function values $F_{AD}\left(\mathbf{U}^{k+1}\right)$ and $F_R\left(\mathbf{U}^{(k+1)}\right)$
    .}

  \BlankLine
  \Comment{Compute integrals}
    $S^{1,k} \leftarrow \sum_{j=0}^M \Delta t q_{1,j} \Big(F_{AD}\left(U^{j,k}\right)+F_{R}\left(U^{j,k}\right)\Big)$ \;
  \For{$m=1 \ldots M-1$}{
    $S^{m+1,k} \leftarrow \sum_{j=0}^M \Delta t \left(q_{m+1,j} - q_{m,j} \right) \Big(F_{AD}\left(U^{j,k}\right)+F_{R}\left(U^{j,k}\right)\Big)$ \;
  }

  \BlankLine
  \Comment{IMEX sub-stepping for correction}
  \For{$m=0 \ldots M-1$}{
    $U^{m+1,k+1} \leftarrow U^{m,k+1} + \Delta t^m \left( F_{AD}^{m,k+1} - F_{AD}^{m,k} 
        +F_R^{m+1,k+1}- F_{R}^{m+1,k} \right) + \Delta t S^{m+1,k}  + \tau^{m,k}$\;
    Evaluate $F_{AD}^{m+1,k+1}$ \; 
    Evaluate $F_{R}^{m+1,k+1} $\; 
  }

  \caption{IMEX SDC sweep algorithm for a single level.}
  \label{alg:sdcsweep}
\end{algorithm}

\begin{algorithm}[H]
  \SetKwComment{Comment}{\# }{}
  \SetCommentSty{textit}
  \DontPrintSemicolon

  \KwData{From previous iteration $k$, initial solution $\mathbf{U}^{k}_{\ell}$ on all levels $\ell$ and function values $F\left(\mathbf{U}^k_{\ell=0}\right)$ on coarse level $\ell=0$.}
  \KwResult{Solution $\mathbf{U}^{k+1}_{\ell}$ and function values $F\left(\mathbf{U}^{k+1}_{\ell}\right)$  on all levels.}

  \BlankLine
  \Comment{Perform SDC sweep on coarse level}
  $\mathbf{U}_{\ell=0}^{k+1}, \mathbf{F}_{\ell=0}^{k+1} \leftarrow {\tt{SDCSweep}}\left(\mathbf{U}_{\ell=0}^{k}, \mathbf{F}_{\ell=0}^{k}\right)$\;

  \BlankLine
  \Comment{Cycle from coarse to fine levels}
  \For{$\ell = 0 \ldots L-1$}{
    \Comment{Interpolate coarse corrections and re-evaluate}
    \For{$m=0 \ldots M$}{
    $U_{\ell+1}^{m,k} \leftarrow {\tt{Interpolate Interior Nodes}}\left(U_{\ell}^{m,k},U_{\ell}^{m,k+1} \right) $ \Comment{\S\ref{par:interior_nodes}} \;
     $U_{\ell+1}^{m,k} \leftarrow {\tt{Interpolate GhostCells}}\left(U_{\ell}^{m=0,k},f \big(U_{\ell}^{m,k+1} \big)\right) $ \Comment{\S\ref{par:ghost_cells}}  \;
     Evaluate $F_{\ell +1}^{m,k} $ \;
   
    }
    $\mathbf{U}_{\ell+1}^{k+1}, \mathbf{F}_{\ell+1}^{k+1} \leftarrow {\tt{SDCSweep}}\left(\mathbf{U}_{\ell+1}^{k}, \mathbf{F}_{\ell+1}^{k},\bm{\tau}^{k}_{\ell+1}\right)$\;  
  
  }

  \BlankLine
  \Comment{Cycle from fine to coarse levels}
  \For{$\ell = L-1 \ldots 0$}{
    \Comment{Restrict, compute FAS and re-evaluate}
    \For{$m=0 \ldots M$}{
    $U_{\ell}^{m,k+1} \leftarrow {\tt{Restrict}}\left(U_{\ell+1}^{m,k+1} \right) $ \Comment{Eq.~(\ref{eqn:restrict_U})} \;
    $\tau^{m,k+1}_{\ell} \leftarrow {\tt{FAS}}\left(F \big(U^{m,k+1}_{\ell} \big), F \big(U^{m,k+1}_{\ell+1} \big) \right) $ \Comment{Eqs~(\ref{eqn:start_FAS}) to (\ref{eqn:end_FAS})}  \;

    }
    $\mathbf{U}_{\ell}^{k+1}, \mathbf{F}_{\ell}^{k+1} \leftarrow {\tt{SDCSweep}}\left(\mathbf{U}_{\ell}^{k+1}, \mathbf{F}_{\ell}^{k+1},\bm{\tau}^{k+1}_{\ell}\right)$\;  
  
  }
  
  \caption{AMLSDC V-cycle iteration for multi-level hierarchy.}
  \label{alg:amlsdc}
\end{algorithm}

\begin{algorithm}[H]
  \SetKwComment{Comment}{\# }{}
  \SetCommentSty{textit}
  \DontPrintSemicolon

  \KwData{Solution $\mathbf{U}^{n}_{\ell}$ on all levels $\ell$ at time $t$ .}
  \KwResult{Solution $\mathbf{U}^{n+1}_{\ell}$ on all levels at time $t + \Delta t$.}

  \BlankLine
  \Comment{Initial solution is spread over all collocation nodes}
  $\mathbf{U}_{\ell}^{k} \leftarrow U_{\ell}^{n}$;
  $\mathbf{F}_{\ell}^{k} \leftarrow f\left(U_{\ell}^{n},t^{m=0} \right)$\;

  \BlankLine
  \Comment{Cycle through AMLSDC iterations}
  \For{$k = 1 \ldots K$}{
    \Comment{Performs V-cycles}
    $\mathbf{U}_{\ell}^{k+1}, \mathbf{F}_{\ell}^{k+1} \leftarrow {\tt{AMLSDC{\_}Vcycle}}\left(\mathbf{U}_{\ell}^{k}, \mathbf{F}_{\ell}^{k}\right)$\;  
    
    \Comment{Check convergence criteria with Eq.(\ref{eqn:convergence_criteria})}    
     \If{ $(k=K)$ {\normalfont \textbf{or}} $(|\mathfrak{R}^{k+1}| < \varepsilon_{\rm SDC})$ }{
      Return
      }
  }

  \BlankLine
  \Comment{Perform last SDC sweep to enforce mass conservation}
  \For{$\ell = 0 \ldots L$}{

    $\mathbf{U}_{\ell}^{n+1}, \mathbf{F}_{\ell}^{n+1} \leftarrow {\tt{SDCSweep}}\left(\mathbf{U}_{\ell}^{k+1}, \mathbf{F}_{\ell}^{k+1},\bm{\tau}^{k+1}_{\ell} \equiv 0 \right)$\;  
  
  }
  
  \caption{Main structure of a time-step iteration from $t^n$ to $t^{n+1}$.}
  \label{alg:whole_amlsdc_RNS}
\end{algorithm}

\section{Results}
\label{sec:results}

In this section we present several numerical tests of the AMLSDC algorithm as implemented in a code referred to
as {\tt RNS}. The results
illustrate the convergence properties of the multi-level temporal integration scheme and the behavior with
adaptively refined spatial grids.
The first three examples are non-reacting flows.  The first of these in Section
\S\ref{subsec:1D_acoustic_wave}
focuses on the temporal convergence properties
of the algorithm for inviscid propagation of an acoustic pulse using a fixed multi-level spatial discretization and demonstrates that no loss of convergence occurs when the pulse passes through coarse/fine grid boundaries.
The second example in \S\ref{subsec:2D_COVO}
demonstrates convergence for viscous simulation of a vortical flow using adaptive meshes that track a moving vortex.
The results demonstrate that the same level of accuracy can be achieved using an adaptive mesh strategy as a uniform
spatial discretization over the entire domain.
The final non-reacting case considers
the roll-up of a Kelvin-Helmholtz instability in \S\ref{subsec:2D_KHI}.
This example, which is highly sensitive to noise, demonstrates that
adaptive mesh refinement does not generate any artifacts at boundaries of coarse and fine grids that impact
the overall solution.
The last four examples consider viscous reacting flows.  The first of these in \S\ref{subsec:1D_Flameball}
examines the temporal convergence
properties of the algorithm on ignition and propagation of a premixed hydrogen flame.  The second example in
\S\ref{subsec:1D_PMF} investigates the accuracy of the flame speed for a premixed methane flame.
This example illustrates the resolution requirements
needed to accurately resolve the flame.  The third reacting case  in
\S\ref{subsec:2D_DME_jet} considers a dimethyl ether jet flame. 
This example provides a more comprehensive validation of the algorithm for reacting flows and demonstrates
the capability of the code to treat models with extremely stiff chemistry.  The final example in \S\ref{subsec:3D_methane_jet}
models a fully three-dimensional turbulent hydrogen jet flame, illustrating the utility of the methodology
for DNS of turbulent reacting flows.


\subsection{Propagation of a Gaussian acoustic pulse}
\label{subsec:1D_acoustic_wave}

The first test case, taken from \cite{mccorquodale:2011}, is the propagation of a Gaussian  acoustic pulse.  The initial conditions are given as
\begin{align}
\rho \left( r \right) &= \left \{ \begin{array}{ll}
\rho_{\rm ref} + A \exp\left(-16 r^2\right)\cos\left(\pi r\right)^6 & \mathrm{if} \; r < 1/2 \\
0 &  \mathrm{if} \; r \ge 1/2, 
\end{array} \right. \\
p\left( r \right) &= p_{\rm ref} + \rho\left( r \right)c_0^2, \\
u_{x,y}\left( r \right) &= 0, \\
E\left( r \right) &= p\left( r \right) / \left(\gamma -1 \right)\rho\left( r \right),
\end{align}
where $A$ is an amplification factor and $r$ is the distance from the center of the domain. Here $\rho_{\rm ref}=1.4$, $p_{\rm ref}=1$, $\gamma=1.4$ and $c_0$ is a reference sound speed defined as $c_0 = \sqrt{\gamma p_{\rm ref}/ \rho_{\rm ref}}=1$.
Note that this test case involves only the Euler equations; the diffusion and reaction terms are set to zero in Eqs.~(\ref{eq:NS:mass}-\ref{eqs:NS:energy}). Moreover, only one fictitious species of unity molecular weight is advanced in the system of equations. 

The propagation of the acoustic pulse will be investigated first in 1D and compared to
the standard fourth-order RK4 algorithm using the WENO spatial reconstruction
described above.
Then, the AMLSDC strategy will be compared to the results reported in \cite{mccorquodale:2011} for 2D and 3D simulations of the propagation of the Gaussian acoustic pulse.

\subsubsection{One-dimensional case}

The domain is periodic with $L_x=1$.
Here the amplification factor is set to $A=1.4 \times 10^{-2}$. We simulate to $t=10$ so that the two acoustic waves travel $10$ times through the computational domain in the left and right direction from the initial pulse, and then merge at the end of the simulation to form the same shape as the initial pulse.

The procedures to perform temporal convergence tests are as follows:
\begin{itemize}
\item simulations are performed over a range CFL numbers: $0.3, 0.5, 0.7, 0.9$ and $1.1$.
\item the computational domain is discretized with three different grid configurations.
Two single-level grids are investigated: a coarse grid, with $N_x=256$ points and a fine grid  $N_x=512$ points.
In addition, a two-level set of grids is also investigated where the computational domain is
fully covered by a coarse grid having $N_x=256$ points with additional level of mesh refinement
on the middle 40\% of the domain.
\end{itemize}

For this initial test we focus on the temporal convergence of the algorithm.
Convergence is measured using the $\mathcal{L}^1$-norm of the difference of the density
between the computed solution and a reference solution defined to be the computed solution for ${\rm CFL}=0.3$:
\begin{equation}
\varepsilon_{\rho} = \mathcal{L}^1_{\rho} \left(S_{sol} - S_{ref} \right) = \frac{{\sum_1^{N_x}}\left|\rho_{sol} - \rho_{ref} \right|}{N_x },
\label{eqn:L1_norm_error}
\end{equation} 
where subscripts {\it{sol}} and {\it{ref}} identify the numerical solution and the reference solution.
Note that for the two-level mesh, a composite error is computed by omitting the solution on parts of the coarse level
that are covered by a finer level of refinement.
We also compare to the a fourth-order AMR Runge-Kutta algorithm coupled to the
fourth-order conservative WENO spatial discretization. 
For these simulations $\varepsilon_{\rm SDC}$ is set to $1 \times 10^{-12}$ and $K=4$.

\begin{figure}[!ht]
\begin{center}
\includegraphics[width=0.75\textwidth]{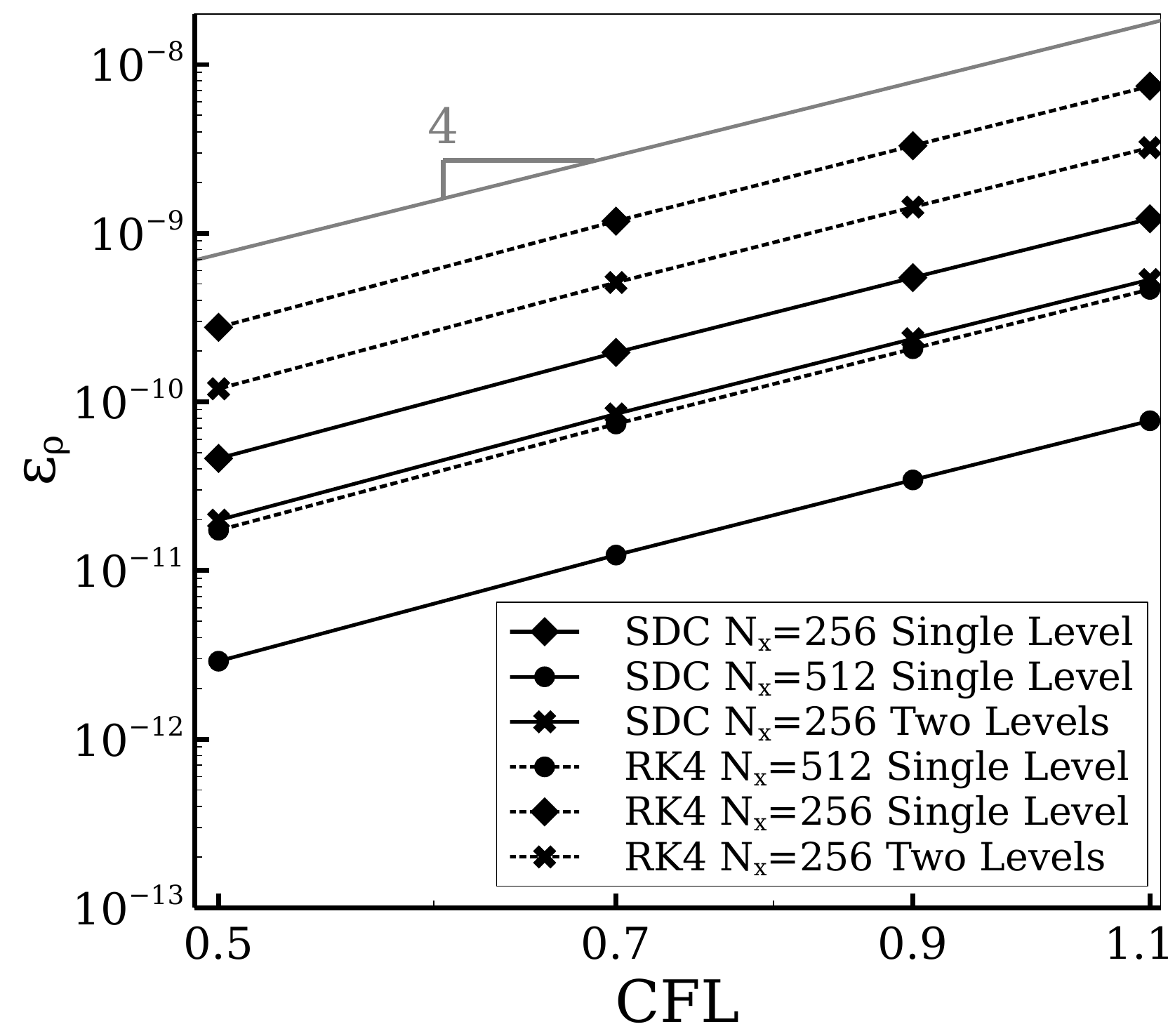}
\end{center}
 \caption{$\mathcal{L}^{1}$-norm of the discretization error $\varepsilon_{\rho}$ in density for different CFL numbers. The diamond and circle symbols represent $\varepsilon_{\rho}$ computed with the single-level mesh discretized with $256$ and $512$ cells, respectively, whereas the cross symbol represents $\varepsilon_{\rho}$ computed with the two-level grids. The solid and dotted lines represent the simulations performed with the AMLSDC and the RK4 methods, respectively. The solid grey line is a slope representing fourth-order convergence.}
 \label{fig:1D_acoustic_wave_time_analysis}
\end{figure}

Results are presented in Figure \ref{fig:1D_acoustic_wave_time_analysis}. The diamond and circle symbols represent $\varepsilon_{\rho}$ computed for different CFL numbers with the single-level mesh discretized with $256$ and $512$ cells, respectively, whereas the cross symbol represents $\varepsilon_{\rho}$ computed for different CFL numbers with the two-level grids. The solid gray line at the top is a slope representing fourth-order convergence. The solid and dotted lines represent the simulations performed with the AMLSDC and the RK4 methods, respectively. 

All simulations demonstrate fourth-order convergence in time. Overall, for all of the different meshes investigated, the discretization errors computed with the AMLSDC method are about a factor of six lower than results
computed with the RK4 method. The solutions computed with the AMLSDC method give a similar numerical error to the solutions computed with the RK4 method, but for a CFL condition number twice as large.
From the results in Figure \ref{fig:1D_acoustic_wave_time_analysis},  we can also evaluate the spatial rate of convergence, which is approximately $3.9$ for each CFL number considered.
As expected, the magnitude of the errors for the two-level simulations lie between those from the coarse and fine uniform grids since
the pulse travels across both coarse and fine grids.
Careful treatment of the coarse / fine boundary for RK4 as described in \cite{mccorquodale:2011} and for
AMLSDC as discussed above avoids any order reduction for the temporal discretization.

\subsubsection{Multi-dimensional cases}

We now consider the multidimensional version of the acoustic wave propagation property
as given in
\cite{mccorquodale:2011}.
In this case, simulations are performed in a periodic domain of dimensions $[0,1]^D$, where $D$ denotes
the dimension.
Here, the amplification factor is set to $A=1.4 \times 10^{-1}$. Other parameters remain the same, except that the
CFL number is fixed at $0.5$. We simulate to $0.24$~on a mesh hierarchy composed of two levels, where coarse grids cover the domain and finer level covers $[1/4,3/4]^D$.

Results are presented  in Tables \ref{tab:acoustic_pulse_2D} and \ref{tab:acoustic_pulse_3D} for the 2D and 3D simulations, respectively. Note that in contrast to the one-dimensional case, the max-norm of the densities between results is reported here. Overall, fifth-order of convergence is observed for the AMLSDC temporal integration strategy coupled with WENO reconstructions implemented in the {\tt RNS} code.
This rate of convergence is higher than expected and suggests that the error is dominated by
WENO reconstruction.
Here we compare directly to the results in 
reported in \cite{mccorquodale:2011}.
For the coarser grids in three dimensions, the algorithm of McCorquodale and Colella is more accurate but at finer
resolution, the AMLSDC is more accurate in both two and three dimensions.  

\begin{table}[!ht]
\centering
\renewcommand\arraystretch{1.3}
\begin{tabular}{c || C{1.5cm} | C{1cm} | C{1.5cm}  | C{1cm}  | C{1.5cm} | C{1cm} | C{1.5cm} }
 & $1/64:1/128$  & rate & $1/128:1/256$ & rate & $1/256:1/512$ & rate & $1/512:1/1024$ \\ \hline\hline 
{\tt RNS}  &  $5.30$e$^{-6}$ & $5.18$ & $1.45$e$^{-7}$ & $5.58$ & $3.04$e$^{-9}$ & $5.06$ & $9.10$e$^{-11}$ \\\hline
 \cite{mccorquodale:2011} & $7.28$e$^{-6}$ & $3.97$ & $4.66$e$^{-7}$ & $3.95$ & $3.01$e$^{-8}$ & $3.99$ & $1.90$e$^{-9}$ \\ 
\end{tabular}
\caption{Convergence of difference in density at time $0.24$ for 2D Gaussian acoustic pulse, computed with fixed grids on two levels with the present {\tt RNS} code (first line) and results reported in \cite{mccorquodale:2011} (second line). Columns alternate between showing the max-norm of the densities between results with the indicated mesh spacing at the coarser of the two levels, and the convergence rate.}
\label{tab:acoustic_pulse_2D}
\end{table}

\begin{table}[!ht]
\centering
\renewcommand\arraystretch{1.3}
\begin{tabular}{c || C{1.5cm} | C{1cm} | C{1.5cm}  | C{1cm}  | C{1.5cm} | C{1cm} | C{1.5cm} }
 & $1/16:1/32$  & rate & $1/32:1/64$ & rate & $1/64:1/128$ & rate & $1/128:1/256$ \\ \hline\hline 
{\tt RNS}  &  $9.35$e$^{-4}$ & $2.93$ & $1.22$e$^{-4}$ & $4.53$ & $5.31$e$^{-6}$ & $5.52$ & $1.16$e$^{-7}$ \\\hline
 \cite{mccorquodale:2011} & $6.84$e$^{-4}$ & $3.39$ & $6.54$e$^{-5}$ & $3.69$ & $5.06$e$^{-6}$ & $3.78$ & $3.70$e$^{-7}$ \\ 
\end{tabular}
\caption{Convergence of difference in density at time $0.24$ for 3D Gaussian acoustic pulse, computed with fixed grids on two levels with the present {\tt RNS} code (first line) and results reported in \cite{mccorquodale:2011} (second line). Columns alternate between showing the max-norm of the densities between results with the indicated mesh spaciness at the coarser of the two levels, and the convergence rate.}
\label{tab:acoustic_pulse_3D}
\end{table}

\subsection{Two-dimensional convection of a diffusive vortex}
\label{subsec:2D_COVO}

The next test case consists of the simulation of the convection of a two-dimensional vortex in a fully periodic square domain including the effects of viscosity.
Here, the formulation of the problem is based on a problem proposed in \cite{Granet:2010}, except the gas mixture is
air ($Y_{{\rm O_2}}=0.233$ and $Y_{{\rm N_2}}=0.767$). The computational domain is a box with $L_x = L_y = 0.01$m. The configuration is a single vortex superimposed on a uniform flow field along both the x and y-directions. The stream function $\Psi$ of the initial vortex is given by

\begin{equation}
\Psi = \Gamma \exp\left(\frac{-r^2}{2 R_v^2} \right), 
\end{equation}
where $r=\sqrt{\left(x-x_0\right)^2 + \left(y-y_0\right)^2}$ is the radial distance from the center of the vortex located at $\left[ x_0, y_0\right]$. Here $\Gamma$ and $R_v$ are the vortex strength and radius, respectively. The velocity field is then defined as

\begin{equation}
u_x = \frac{\partial  \Psi}{\partial x} + u_{y, \rm ref}, \hspace{1cm} u_y = \frac{\partial  \Psi}{\partial y} + u_{y,\rm ref}.
\end{equation}
The initial pressure field is given by
\begin{equation}
p\left(r\right) = p_{\rm ref} \exp\left(-\frac{\gamma}{2} \left(\frac{\Gamma}{c R_v} \right)^2 \exp \left( -\frac{r^2}{R_v^2} \right) \right),
\end{equation}
and the corresponding density and energy fields are computed through the equation of state, assuming a constant temperature $T_{\rm ref}=300$K. 
The pressure $p_{\rm ref}$ is set to $101325$Pa. The vortex is located at $\left[ x_0, y_0\right]$ = $\left[ 0, 0\right]$
with parameters $\Gamma = 0.11$m$^2$/s and $R_v = 0.1$.
The vortex is convected in the diagonal direction at a uniform velocity $u_{\rm ref}=(10,10)$m/s.
The simulations are performed over a physical time of $5$ms, corresponding to 5 flow through times (FTT), in order to accumulate enough numerical errors from the spatial discretization schemes. Note that the Reynolds number is about $Re=2100$.
At this Reynolds number, the viscous effects are significant so that after 5 FTT we can measure the effects of both
advection and diffusion.

Two different sets of meshes are tested. First, simulations are performed on single level uniform grids
with  $N_x=N_y= 16, 32, 64, 128, 256, 512$.
Second, three sets of two-level grids are used: the coarse grid are chosen to be   $N_x,N_y= 16, 32, 64, 128, 256$, and an additional level of mesh refinement is super-imposed over the vortex. In order to make this patch of mesh refinement follow the vortex during its convection, a cell tagging criterion based on the absolute value of the vorticity $\left( \omega=\frac{\partial u_y}{\partial x}-\frac{\partial u_x}{\partial y}\right)$ is used. Basically, all cells on the coarse grid where $|\omega| > 1$ are tagged for refinement. Similar to the study performed in \S\ref{subsec:1D_acoustic_wave} for the one-dimensional acoustic wave, the $\mathcal{L}^1$-norm errors $\varepsilon$ for the $x-$velocity are computed using Eq.~(\ref{eqn:L1_norm_error}). Here, the reference solution is chosen to be the next finest solution. Furthermore, all simulations are performed with a fixed  ${\rm CFL}=0.5$ and $\varepsilon_{\rm SDC}$ is set to $1 \times 10^{-12}$, with $K=4$.

\begin{figure}[ht]
\begin{center}
\includegraphics[width=0.75\textwidth]{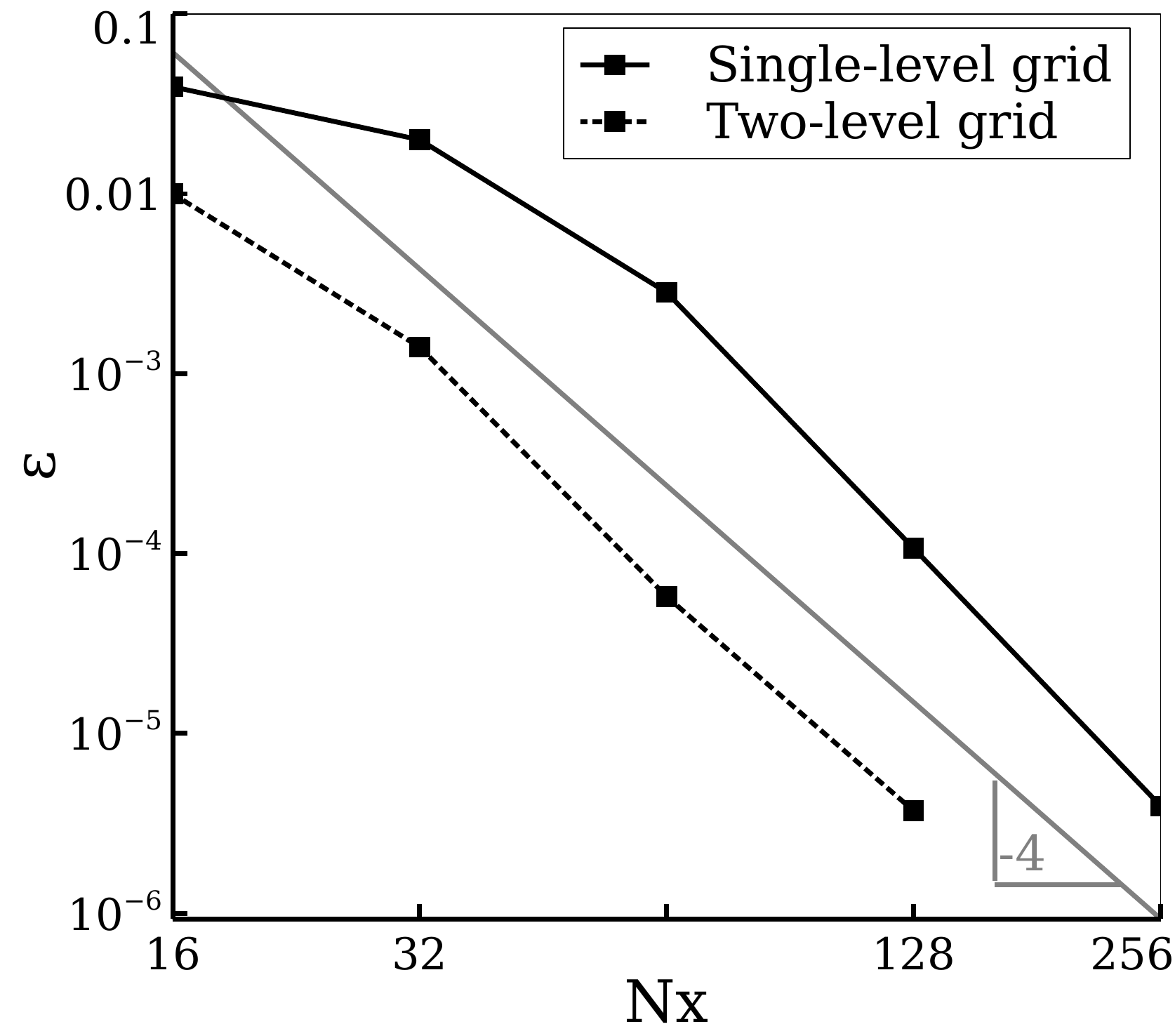}
\end{center}
 \caption{$\mathcal{L}^{1}$-norm of the discretization error $\varepsilon$ computed for the x-velocity and for different mesh discretizations. The black solid and dotted lines represent $\varepsilon$ computed with the single-level and two-level grid configurations, respectively. The gray solid line is a virtual slope representing fourth-order convergence.}
 \label{fig:2D_COVO_space_analysis}
\end{figure}

Results are presented in Figure~\ref{fig:2D_COVO_space_analysis}. The black solid and dotted lines represent the $\mathcal{L}^1$-norm errors $\varepsilon$ computed for the single-level and two-level simulations, respectively.
The gray solid line represents fourth-order convergence.
Both the uniform grid and the adaptive grid simulations show
fourth-order convergence, consistent with the order of accuracy
of the discretization.
Furthermore, the errors for the adaptive simulations are comparable to the uniform grid at the same resolution
as the finest level in the hierarchy, demonstrating that the dynamic refinement is accurately tracking the important
parts of the flow field.

\subsection{Two-dimensional Kelvin-Helmholtz Instability}
\label{subsec:2D_KHI}

In \S\ref{subsec:1D_acoustic_wave} and \S\ref{subsec:2D_COVO} we showed  that the AMLSDC strategy
coupled with the conservative finite-volume  method exhibited fourth-order in both space and time,
even with dynamic mesh refinement.
The next test case investigates a more complex and challenging problem.
As discussed in \cite{Lecoanet:2016}, the Kelvin-Helmholtz instability problem is very sensitive to numerical discretization errors due to the highly nonlinear behavior of this type of flow.
It has been shown that numerical errors can seed spurious nonphysical small-scale structures in the flow. Moreover, the authors in \cite{Lecoanet:2016} formulated initial conditions to ensure that no spurious numerical errors result from boundary conditions or from randomly generated noise that can trigger the instability mechanism.

The test case consists
a central layer moving to the right with flow above and below moving to the left. A
vertical velocity perturbation is super-imposed to initiate the instability.
The initial conditions are given as follows:
\begin{align}
\rho\left(x,y\right) &= 1, \\
u_x \left(x,y\right) &= u_{\rm ref} \times \left[ \tanh \left(\frac{y-y_1}{a}   \right)  - \tanh \left(\frac{y-y_2}{a} \right)  -1 \right], \\
u_y \left(x,y\right) &= A \sin \left(2 \pi x \right) \times \left[ \exp \left(-\frac{(y-y_1)^2}{\sigma^2} \right) + \exp \left(-\frac{(y-y_2)^2}{\sigma^2} \right) \right], \\
p \left(x,y\right) &= p_{\rm ref}, \\
Y \left(x,y\right) &= \frac{1}{2} \left[ \tanh \left(\frac{y-y_2}{a} \right) -  \tanh \left(\frac{y-y_1}{a} \right) +2 \right]. \label{eqn:KHI_species_Y1} 
\end{align}
Here, $a=0.05$, $\sigma=0.2$ and $A=0.01$ are parameters controlling the flow structure and the velocity perturbation, while $\gamma=5/3$, $u_{\rm ref}=10$m/s and $p_{\rm ref}=10$Pa, so that the overall Mach number is $M \approx 0.25$. The computational domain is a rectangular domain with $L_x = 1$m and $L_y=2$m.
We also set $y_1=0.5$ and $y_2=1.5$.
Here
a non-diffusive tracer is introduced to track the mixing process of the two fluid layers.

First, simulations are performed for single-level grids of resolutions $128 \times 256$, $256 \times 512$, $512 \times 1024$ and $1024 \times 2048$ cells. Second, AMR simulations are performed by discretizing the coarsest level with $128 \times 256$ cells, and successively adding levels of mesh refinement with
a refinement ratio of $2$.
The refinement criterion is based on the gradient of the fictitious species, and coarse cells are refined if $|\nabla Y|/Y > 0.9\Delta x$. In all simulations, the time-step is set to $\Delta t=1 \times 10^{-4}$s.
Note that in order to make a more sensitive test, this case uses the Euler equations; diffusion and reactive terms are set to zero in Eqs.~(\ref{eq:NS:mass}-\ref{eqs:NS:energy}). Note that $\varepsilon_{\rm SDC}$ is set to $1 \times 10^{-12}$ and $K=4$.

\begin{figure}[h]
\begin{center}
\includegraphics[width=0.75\textwidth]{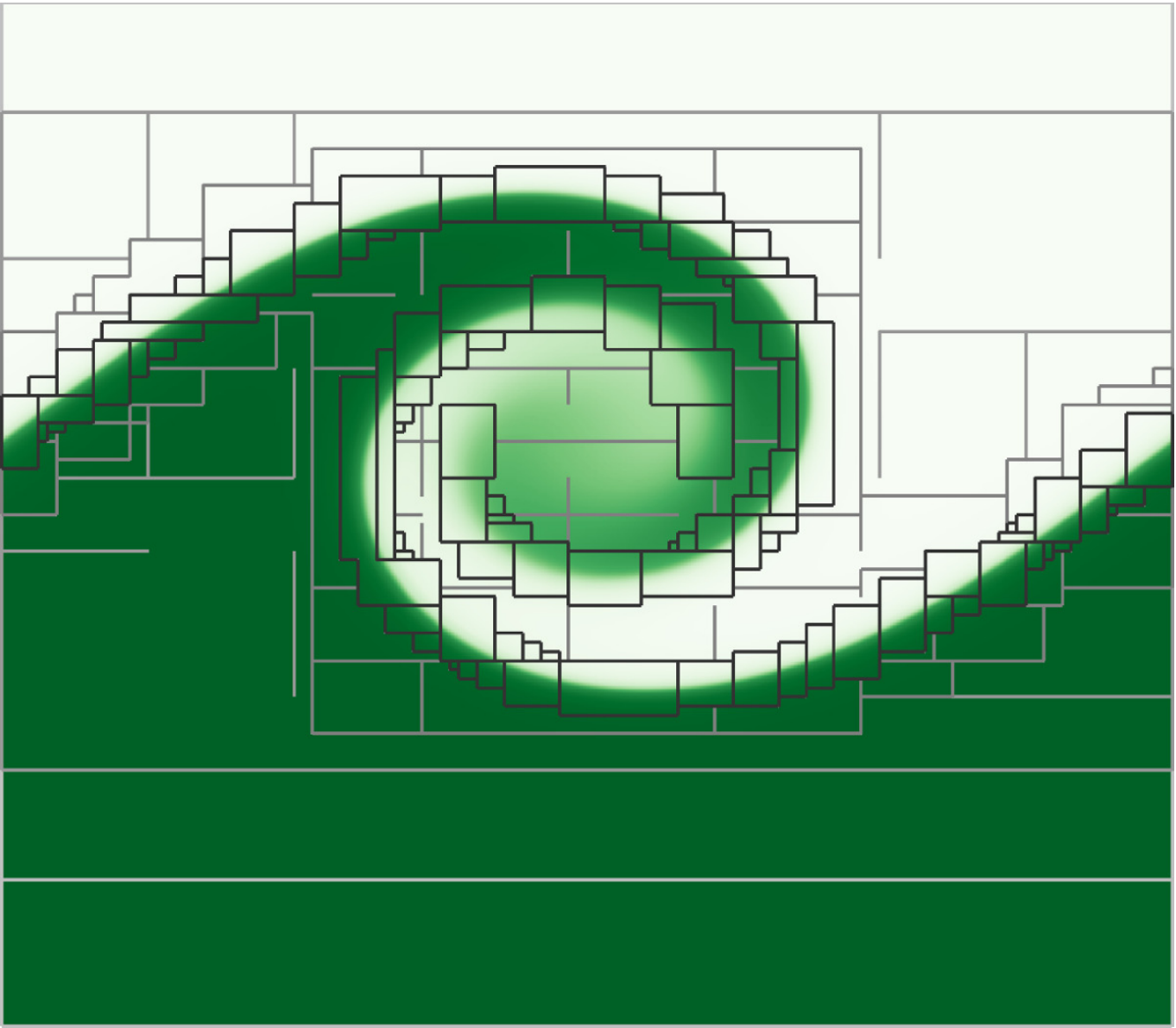}
\end{center}
 \caption{Field of the fictive species $Y$ at time $t=2$s for $4$ levels of mesh refinement. The color map ranges from $Y=0$ (white color) to $Y=1$ (green color).  White, gray and black boxes represent the domain decomposition on the second, third and fourth level of mesh refinement, respectively. Note that the image shows only the low portion of the domain.}
 \label{fig:2D_KHI_wavy}
\end{figure}

Figure~\ref{fig:2D_KHI_wavy} presents the field of fictitious species $Y$ at time $t=2$s for $4$ levels of mesh refinement. The color map ranges from $Y=0$ (white color) to $Y=1$ (green color). The Kelvin-Helmholtz instability process is clearly observable, with the two layers rolling over each other to create a vortex.
Figure~\ref{fig:2D_KHI_wavy} shows the decomposition of the domain for each level of mesh refinement. White, gray and black boxes mark the second, third and fourth level of mesh refinement, respectively. As expected, the finest level of mesh refinement follows the interface between the flow mixing layers.

\begin{figure}[!ht]
 \begin{subfigmatrix}{1}
  \subfigure[]{\includegraphics[width=0.72\textwidth]{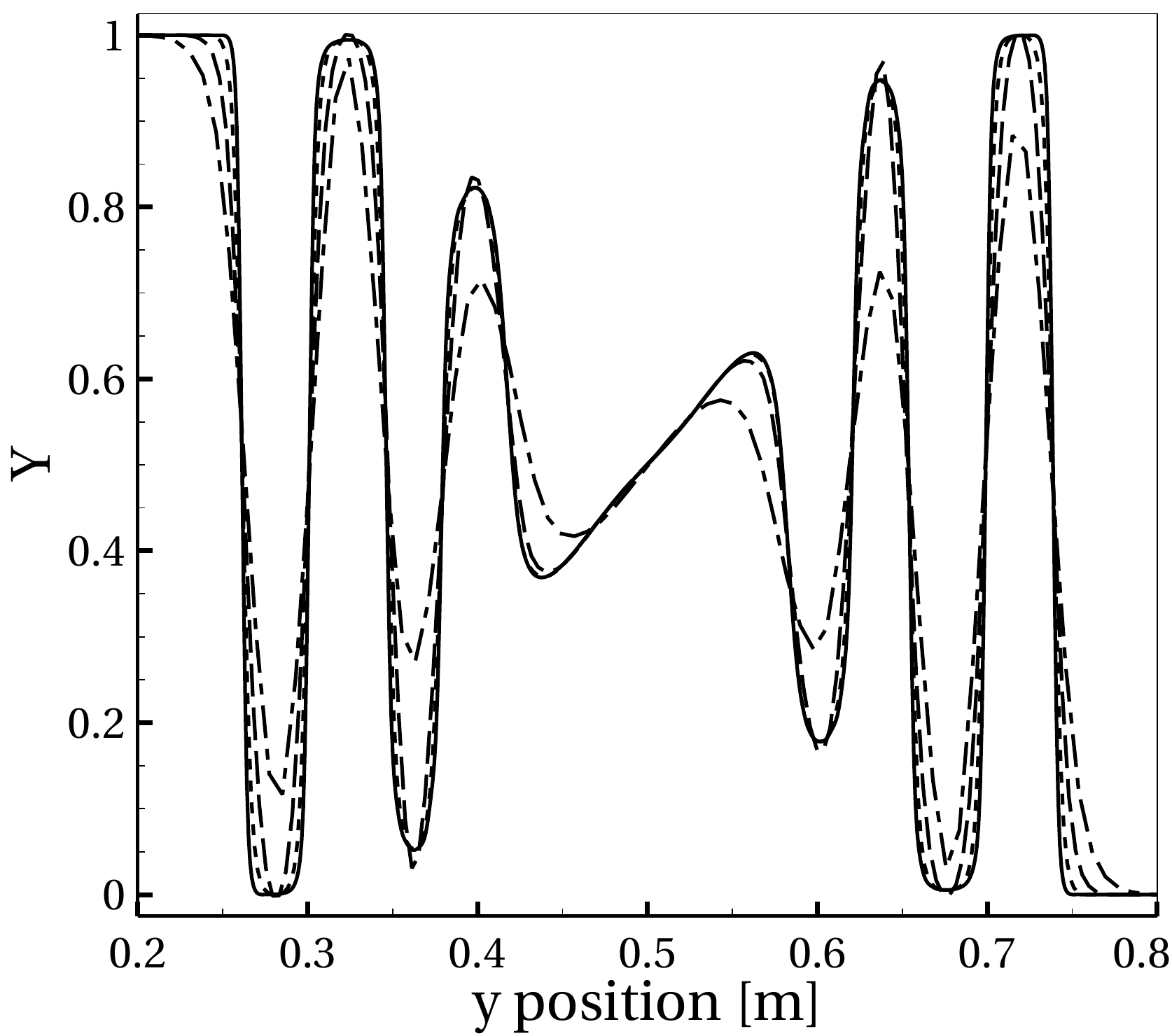}}
  \subfigure[]{\includegraphics[width=0.72\textwidth]{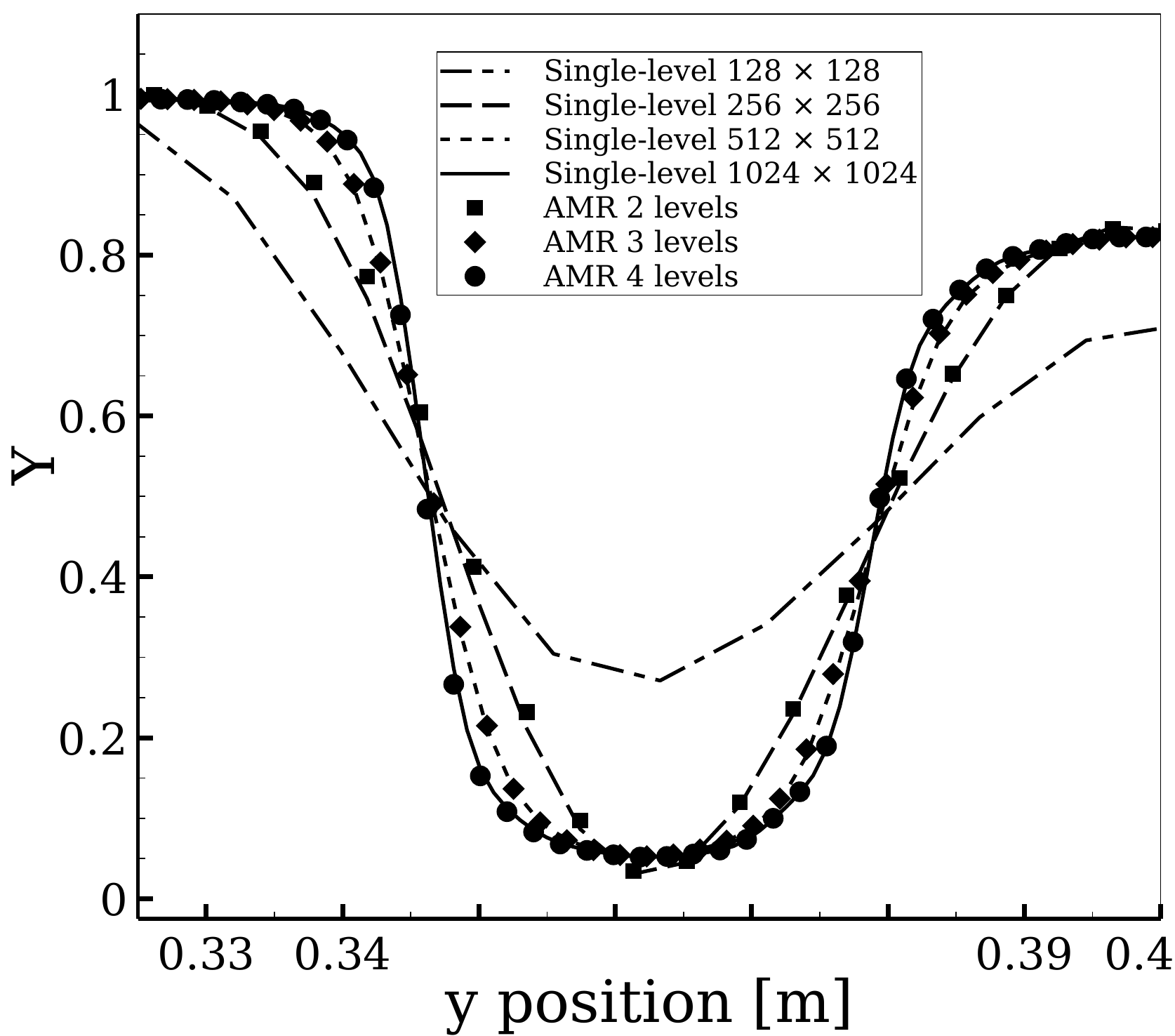}}
  \end{subfigmatrix}
 \caption{Profile of the fictive species $Y$ taken along the $y-$axis at the location $x=L_x/2$, and for different uniform mesh resolutions. The dashed-dotted, dashed, dotted and solid lines correspond to simulations performed on a uniform unique level composed of $128 \times 256$, $256 \times 512$, $512 \times 1024$ and $1024 \times 2048$ points, respectively. The square, diamond and circle symbols represent AMR simulations performed with $2$, $3$ and $4$ levels of mesh refinement. Panel.(b) is a zoom of Panel.(a).}
 \label{fig:2D_KHI_quantitative}
\end{figure}

Quantitative results are presented in Figure~\ref{fig:2D_KHI_quantitative}a and~\ref{fig:2D_KHI_quantitative}b.
Figure~\ref{fig:2D_KHI_quantitative}a shows the profile of the fictitious species $Y$ taken along the $y-$axis at
$x=L_x/2$ for different uniform mesh resolutions. The dashed-dotted, dashed, dotted
and solid lines correspond to simulations performed on uniform grids of $128 \times 256$, $256 \times 512$, $512 \times 1024$ and $1024 \times 2048$ points, respectively. For clarity, only a portion of the domain is presented. In this figure, the convergence of the solution
under mesh refinement is clearly observable.
Figure~\ref{fig:2D_KHI_quantitative}b shows a zoom of Figure~\ref{fig:2D_KHI_quantitative}a, with the addition of results from the AMR simulations. The square, diamond and circle symbols represent AMR simulations performed with $2$, $3$ and $4$ levels of mesh refinement, respectively. 
The AMR simulations achieve roughly the same accuracy as the corresponding uniform grid at the resolution of the finest
AMR level.  This demonstrates that the AMR algorithm can accurately represent the solution without introducing
artifacts at the coarse / fine boundaries that impact the flow.


\subsection{One-dimensional ignition and propagation of a premixed hydrogen flame}
\label{subsec:1D_Flameball}

Now we address the behavior of the AMLSDC algorithm for reacting flows.
The first test case investigates the convergence properties of the AMLSDC strategy with detailed chemical kinetics.
The physical problem is based on the test case in \cite{Emmett:2014}. It consists of the ignition and propagation of a one-dimensional premixed hydrogen flame. A $9-$species H$_2$/O$_2$ reaction mechanism \cite{LiDryer:2004} is used.
The computational domain is $ (-2$mm$, 2$mm$)$ with periodic boundaries. The initial pressure, temperature and velocity of the flow are set to
\begin{align}
p(x) = p_0 \left[1+ 0.1 \exp \left(-\frac{x^2}{r_0^2} \right) \right], \\
T(x) = T_0 + T_1 \exp \left(-\frac{x^2}{r_0^2} \right), \\
u(x) = u_0  \sin\left(\frac{2 \pi}{L_x}x \right),
\end{align}
where $p_0=1$atm, $T_0=300$K, $T_1=700$K, $u_0=3$m.s$^{-1}$ and $r_0=0.1$mm.
Moreover, the mole fractions of species composing the mixture are initially set to zero, except that
\begin{align}
X({\rm H}_2) = 0.5 + 0.025 \exp \left(-\frac{x^2}{r_0^2} \right), \\
X({\rm O}_2) = 0.25 + 0.05 \exp \left(-\frac{x^2}{r_0^2} \right), \\
X({\rm N}_2) = 1 - X({\rm H}_2) - X({\rm O}_2).
\end{align}

For this case we focus on temporal accuracy.
The simulations are performed over a physical time of $8 \times 10^{-6}$s, with $\varepsilon_{\rm SDC}$  set to $1 \times 10^{-12}$ and $K=4$. The procedures to perform the temporal convergence tests are as follows:
\begin{itemize}
\item simulations are performed with time-steps: $\Delta t=2,2.5,3.75,5,6.25,7.5,10 \times 10^{-9}$s,
\item the computational domain is discretized with three different grid configurations. First we consider a baseline mesh with
$N_x=64$ points. The other two configurations are obtained by adding one or two levels of mesh refinement over the entire domain. 
In this case the coarser grids provide an initial approximation to the finer grids, similar to multigrid algorithms.
\end{itemize}

\begin{figure}[!ht]
\begin{center}
\includegraphics[width=0.75\textwidth]{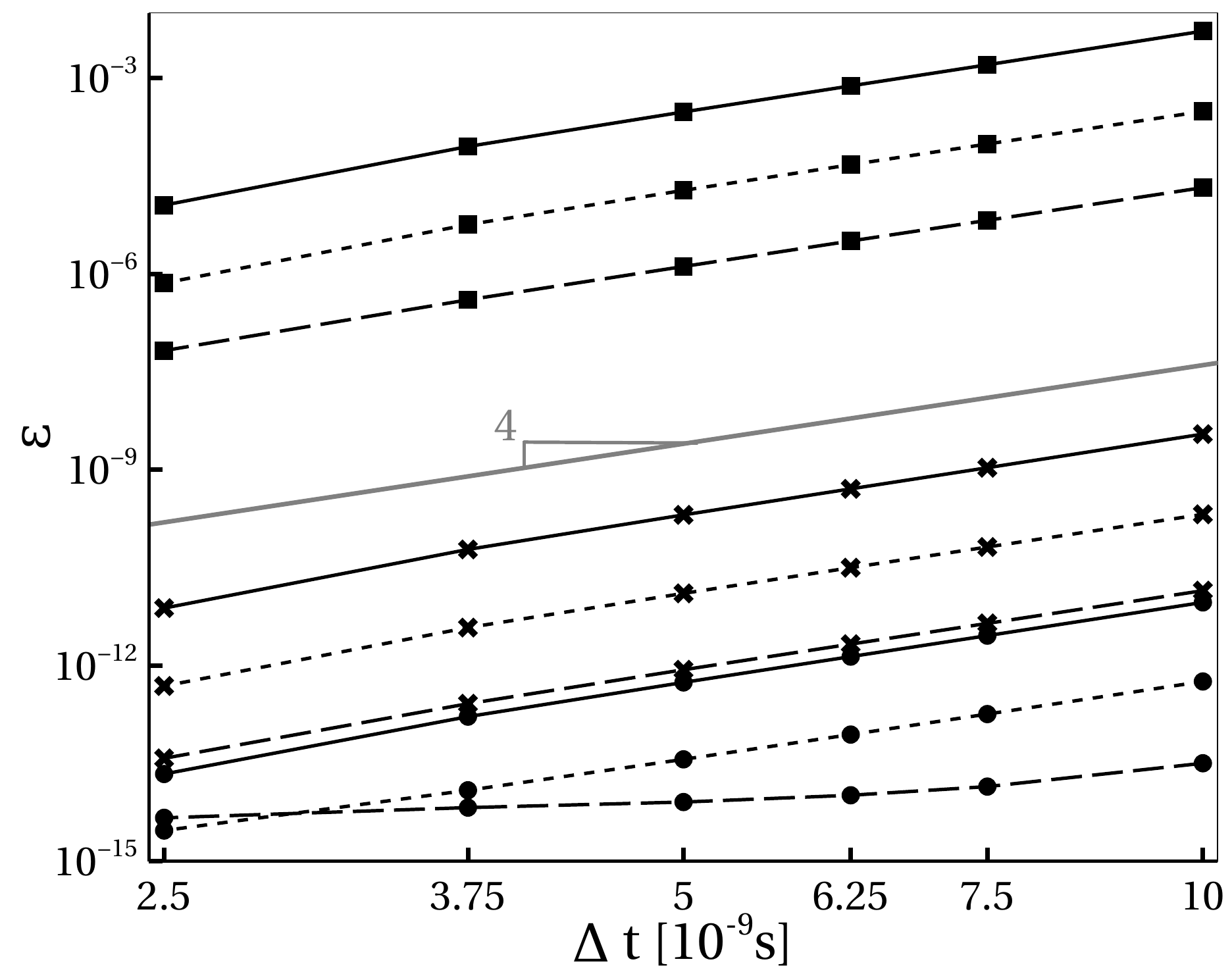}
\end{center}
 \caption{$\mathcal{L}^{1}$-norm of the discretization error $\varepsilon$ computed for $X({\rm HO}_2)$, $X({\rm N}_2)$ and the velocity $u$, represented by the circle, square and cross symbols, respectively. The solid, dotted and dashed lines represent simulations computed with the uniform mesh ($N_x=64$), the two-level and three-level AMR meshes, respectively. The gray solid line is a virtual slope representing fourth-order convergence.}
 \label{fig:1D_Flameball_convergence}
\end{figure}

Figure~\ref{fig:1D_Flameball_convergence} presents the $\mathcal{L}^1$-norm errors $\varepsilon$ for a selection of variables.
Here, the reference solution is the solution with $\Delta t = 2 \times 10^{-9}$s. In Figure~\ref{fig:1D_Flameball_convergence}, the circle, square and cross symbols represent the variables $X({\rm HO}_2)$, $X({\rm N}_2)$ and the velocity $u$, respectively. The solid, dotted and dashed lines represent simulations computed with the uniform mesh ($N_x=64$) and the two-level and three-level meshes, respectively.
Fourth-order temporal convergence is observed
except for $X({\rm HO}_2)$ at the finest level of mesh refinement.
Because of the relatively low mole fraction of this species,
the numerical error is close to machine precision and round-off errors pollute the result.

The temporal rates of convergence have been computed from the $\mathcal{L}^1$-norm errors $\varepsilon$ by linear regression,
and results are reported in Table~\ref{tab:1D_Flameball_convergence} for a selection of variables.
Again, an overall fourth-order convergence rate is observed for all the variables and for all set of mesh grids, except for $X({\rm HO}_2)$ because the errors become too small to capture the actual rate of convergence of the AMLSDC method (see discussion above).
From the results in Figure \ref{fig:1D_Flameball_convergence} we can also the estimate the spatial convergence rate, which is between $3.8$ and $4.0$ depending on the variables investigated.

\begin{table}[!ht]
\centering
\renewcommand\arraystretch{1.3}
\begin{tabular}{c || c | c | c   }
& \multicolumn{3}{c}{Mesh} \\
Variable    & $1$ level  & $2$ levels & $3$ levels    \\ \hline\hline 
$\rho$  & $4.386$ & $4.332$ & $3.97$   \\\hline
$T$  & $4.383$ & $4.332$ & $4.0$   \\\hline  
$u$  & $4.386$ & $4.329$ & $4.133$   \\\hline  
$X({\rm HO}_2)$  & $4.329$ & $3.802$ & $1.28$   \\\hline 
$X({\rm H}_2 {\rm O}_2)$  & $4.384$ & $4.33$ & $4.246$   \\\hline 
$X({\rm N}_2)$  & $4.386$ & $4.329$ & $4.24$   
\end{tabular}
\caption{Rates of convergence computed from the $\mathcal{L}^1$-norm errors $\varepsilon$ by using a best-fitting method.}
\label{tab:1D_Flameball_convergence}
\end{table}

\subsection{One-dimensional premixed methane/air flame}
\label{subsec:1D_PMF}

The next test case assesses the performance of the AMLSDC strategy with AMR,
for computing the laminar flame speed of a one-dimensional premixed methane/air mixture.  Here,
the flame speed $S_L$ is computed from the fuel consumption using
\begin{equation}
S_L = \frac{\int_0^{L} \dot{\omega}_{{\rm CH}_4}}{\rho^u Y_{{\rm CH}_4}^u}
\label{eqn:flame_speed_SL}
\end{equation}
where $L$ is the length of the computational domain and the superscript $u$ corresponds to values in the unburnt region.

In order to provide initial data for the {\tt RNS} code,
the premixed methane/air flame is first computed with the low-Mach number solver {\tt PeleLM} \cite{Nonaka:2018}.
The low Mach number approach allows larger time steps, leading to faster convergence
in cases where the low Mach number assumption is justified
\cite{Majda:1985,Day:2000,Motheau:2016a}.
Moreover, the {\tt PeleLM} software employs an active control strategy \cite{Bell:2006} to adapt the inflow velocity so that the flame will stabilize at a desired location in the computational domain.
Methane and air are injected at the inflow of a 1D domain with a mixture ratio $\phi=0.7$. The corresponding mass fractions imposed at the inlet are $Y_{{\rm CH}_{4}}=0.03926$, $Y_{{\rm O}_{2}}=0.22374$ and $Y_{{\rm N}_{2}}=0.737$, while other species are set to zero. At the opposite end of the domain, a characteristic outflow condition is imposed.
The flow is initialized with a low resolution approximation to the flame.
We model methane combustion using the {\tt GRIMECH 3.0} mechanism \cite{GRIMECH}, which consist of $53$ species and $325$ chemical reactions.
At convergence of {\tt PeleLM}, the flame is stabilized in the middle of the computational domain and the inflow velocity corresponds to the actual laminar flame speed $S_L$. Here $S_L=18.89$cm/s, which is in agreement with published measurements and computational studies \cite{Maaren:1994}. The solution is then extracted as an initial solution for the {\tt RNS} code.

Several simulations are performed with the {\tt RNS} code. The length of the computational domain is $L_x=0.03$m. 
The domain is discretized with a base mesh with $N_x=128$ points, corresponding to
$\Delta x \approx 0.23$mm, which gives approximately five points across the flame fronts. The refinement criterion for AMR is set
so that cells where $Y_{\rm HCCO} > 5 \times 10^{-7}$ are tagged for mesh refinement up to a maximum of $3$ additional levels.
Thus, at the finest level of mesh refinement $\Delta x \approx 0.03$mm, which corresponds to a flame front discretized with about $40$ points. Note that $\varepsilon_{\rm SDC}$ is set to $1 \times 10^{-12}$ and $K=4$; however, in practice convergence is reached for two AMLSDC iteration ($K=2$). 

The case with 3 levels of refinement bears further scrutiny.
As explained in \S\ref{subsec:MLSDC}, the number of Gauss-Lobatto (GL) collocation nodes are increased on each additional level of mesh refinement.  The baseline algorithm uses $3$ GL nodes, so that for 3 levels of refinement,
the finest level uses $17$ GL nodes.
In order to reduce the number of GL collocation nodes at the finest level, another strategy is investigated here.
The maximum number of GL collocation nodes is set to $9$ so that the two first levels of the mesh grid hierarchy use
the same time step (no subcycling). Thus both level 0 and level 1 only use $3$ GL nodes.

\begin{figure}[!ht]
 \begin{subfigmatrix}{2}
  \subfigure[]{\includegraphics[height=0.9\textwidth]{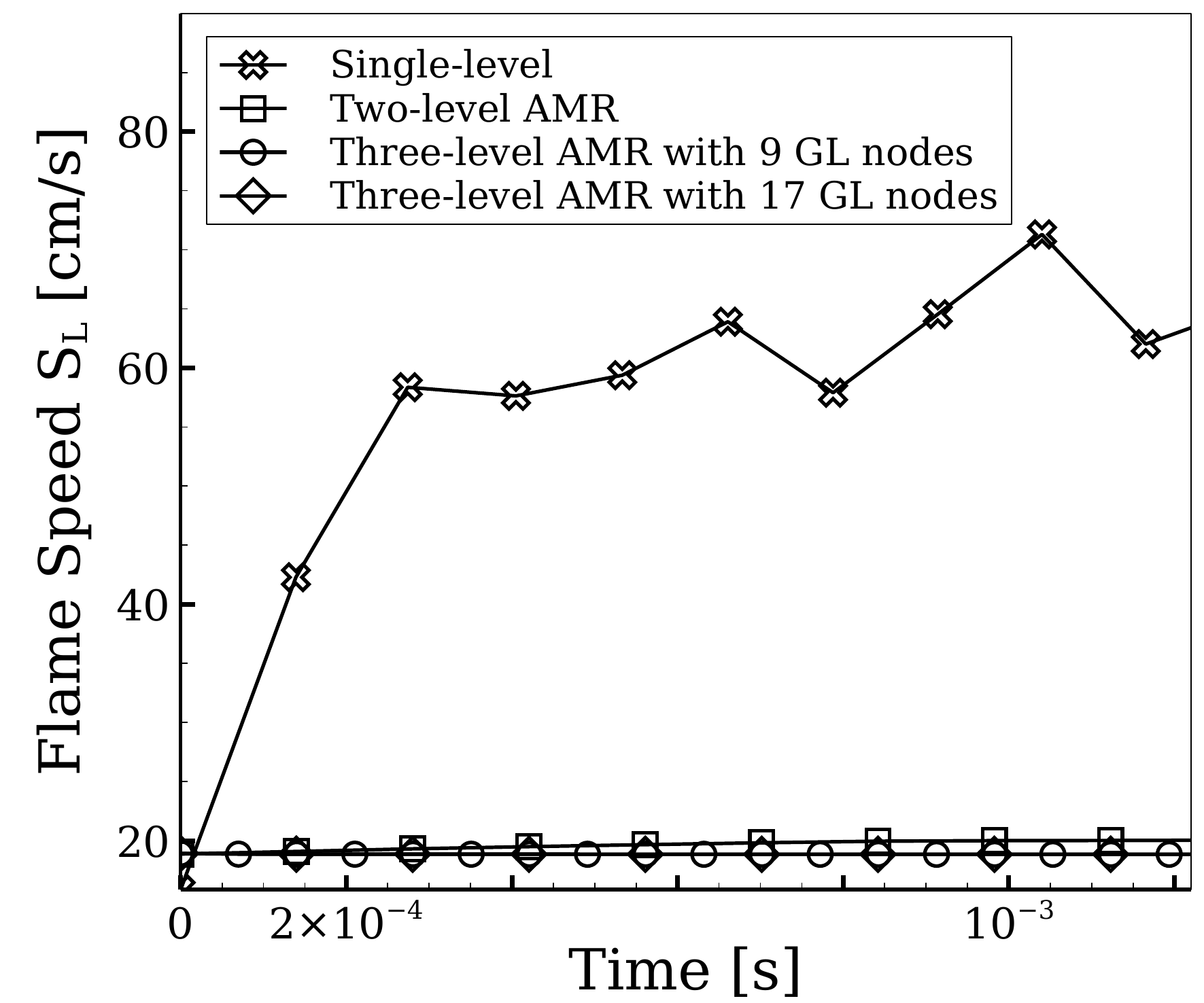}} 
  \subfigure[]{\includegraphics[height=0.9\textwidth]{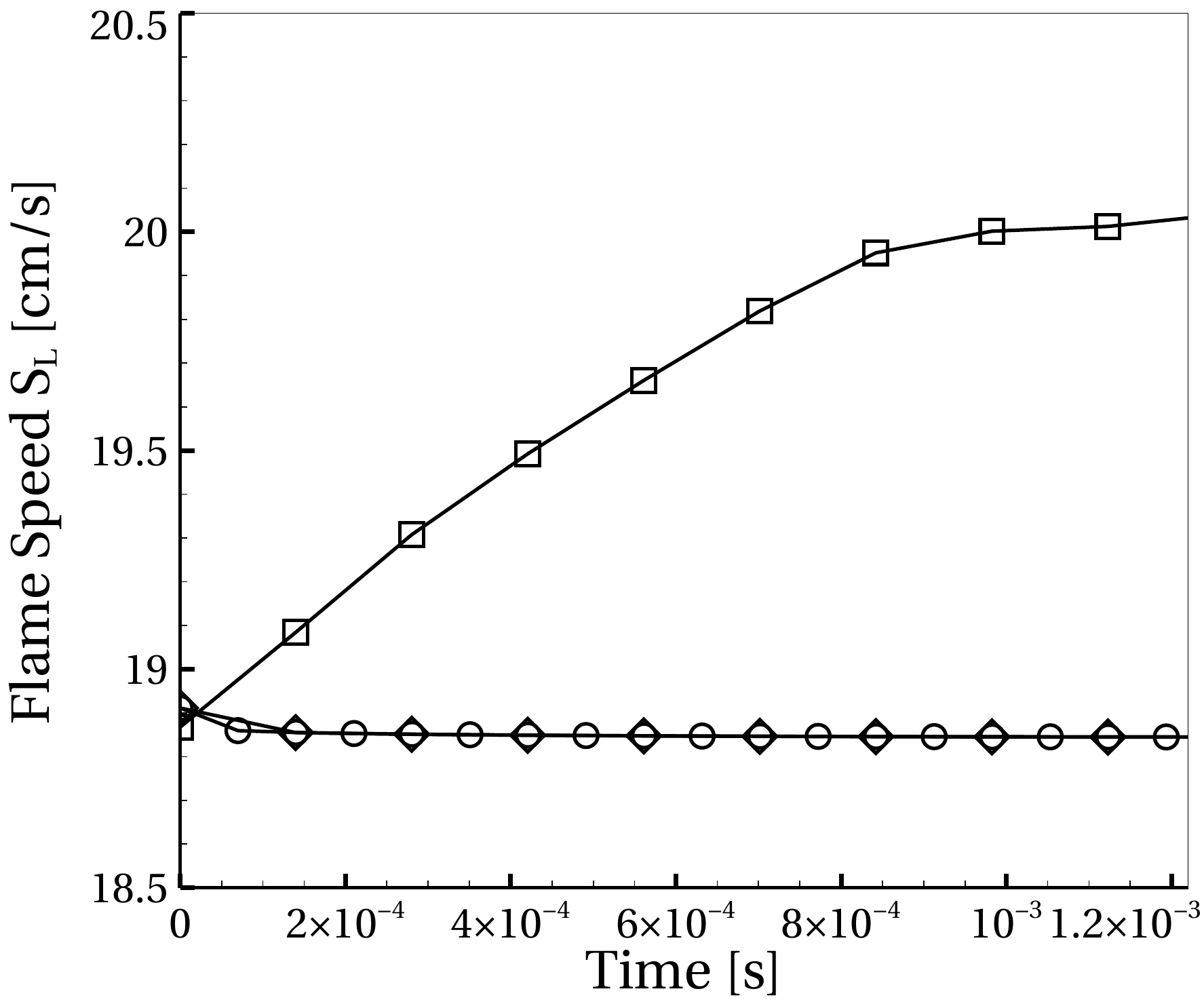}}
  \end{subfigmatrix}
 \caption{Time evolution of the flame speed computed for different levels of mesh refinement. The cross and square symbols represent the solution computed with $1$ and $2$ levels of mesh refinement, respectively. Circle and diamond symbols represent the solution computed with $3$ levels of mesh refinement and with $9$ and $17$ Gauss-Lobatto collocation nodes, respectively. Panel b is a zoom of Panel a.}
 \label{fig:1D_PMF_flame_speed}
\end{figure}

\begin{figure}[!ht]
\begin{center}
\includegraphics[width=0.75\textwidth]{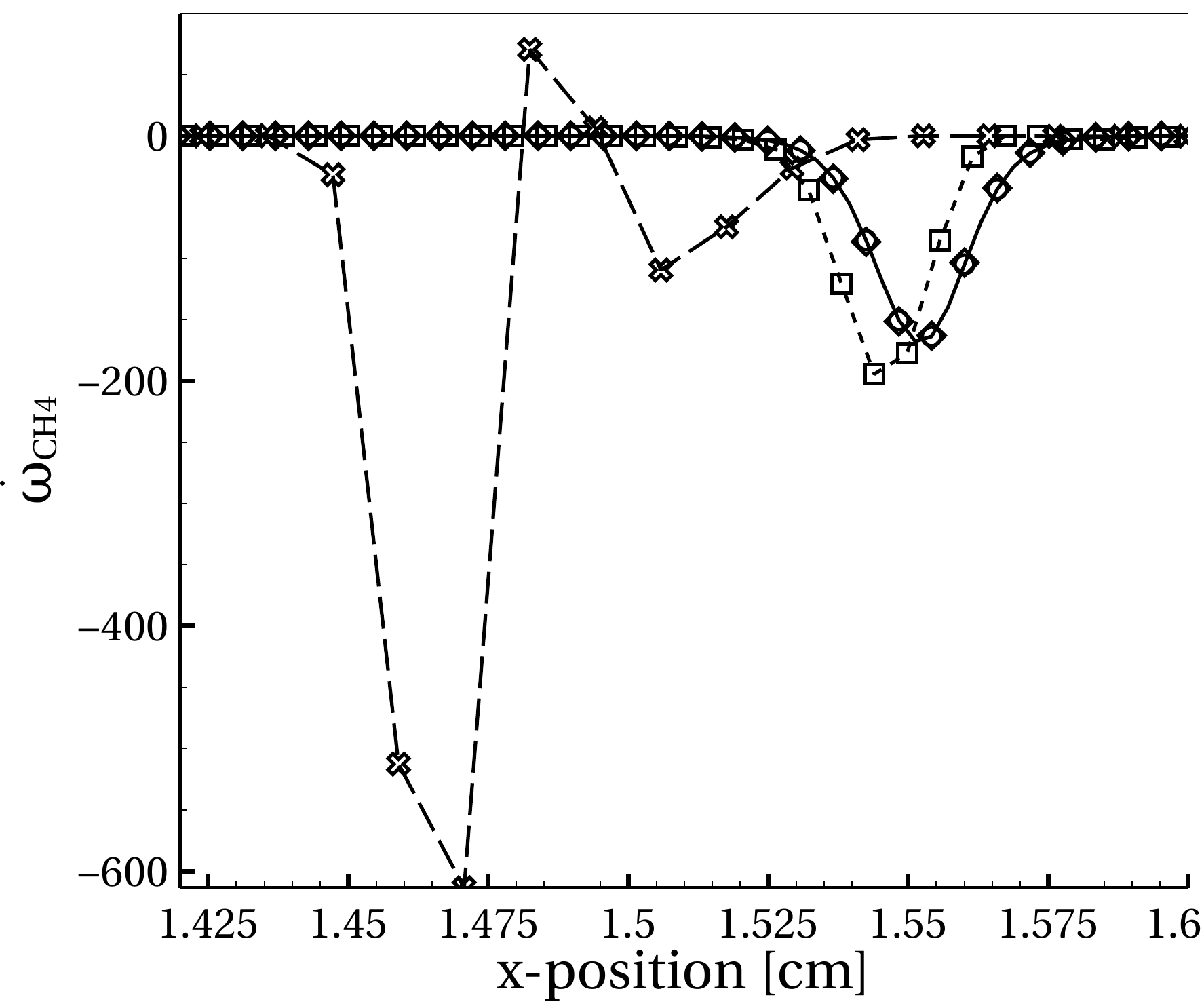}
\end{center}
 \caption{Profile of $\dot{\omega}_{{\rm CH}_4}$ in the flame front at the physical time $1.2 \times 10^{-3}$s. The cross and square symbols represent the solution computed with $1$ and $2$ levels of mesh refinement, respectively. Circle and diamond symbols represent the solution computed with $3$ levels of mesh refinement and with $9$ and $17$ Gauss-Lobatto collocation nodes, respectively.}
 \label{fig:1D_PMF_reaction_rate_CH4}
\end{figure}

The time evolution of the flame speed computed for different levels of mesh refinement is shown in Figure~\ref{fig:1D_PMF_flame_speed}. Note that  Figure~\ref{fig:1D_PMF_flame_speed}b is a zoom of Figure~\ref{fig:1D_PMF_flame_speed}a.
The cross and square symbols represent the solution computed with $1$ and $2$ levels of mesh refinement, respectively.
Circle and diamond symbols represent the solution computed with $3$ levels of mesh refinement using $9$ and $17$ Gauss-Lobatto collocation nodes, respectively. Figure~\ref{fig:1D_PMF_reaction_rate_CH4} presents the profile of $\dot{\omega}_{{\rm CH}_4}$ in the flame front at physical time $1.2 \times 10^{-3}$s. The symbols are the same as in Fig.~\ref{fig:1D_PMF_flame_speed}.

Results for the simulation with $1$ additional level of mesh refinement are obviously not physical. For the simulation with $2$ additional levels of mesh refinement, the flame speed stabilizes around $20$cm/s after a long simulation time, and the profile of $\dot{\omega}_{{\rm CH}_4}$ no longer shows nonphysical oscillations. When $3$ additional levels of mesh refinement are imposed, the flame speed quickly reaches a stable value of $S_L=18.85$cm/s, which is very close to the results from the low-Mach-number simulation. This study shows the impact of the mesh resolution on the ability of the numerical methods to accurately capture the combustion process, and demonstrates that the AMR strategy helps to capture the physics by only refining the mesh in the region of interest in order to save computational effort. 

Table~\ref{tab:1D_PMF_delta_T} presents the coarse time-step $\Delta t$ and
the mean wall-clock CPU time for a coarse time step for each simulation performed.
When each level is refined in time as well as space, the algorithm takes the same coarse time step.
However when the second level is not refined in time compared to level zero,
the minimum time-step is divided by a factor of $2$ in order to maintain the stability of the algorithm. In this case, the mean wall-clock CPU time per step is lower than when $17$ Gauss-Lobatto
nodes are use at the finest level.
However, since the global time-step is also lower by a factor of 2,
the overall computational time is consequently larger. From results shown in Figs.~\ref{fig:1D_PMF_flame_speed} and~\ref{fig:1D_PMF_reaction_rate_CH4}, both strategies give virtually identical results.
While reducing the maximum number of Gauss-Lobatto collocation nodes may not seem to be efficient
in the present one-dimensional test case, we emphasize that a large number of GL collocation nodes at the finest level will
lead to large memory requirements three-dimensional simulations.
This requirement will become worse if additional levels of mesh refinement are used,
which highlights an important trade-off between memory usage and 
computational efficiency of the AMLSDC strategy presented in this paper.

\begin{table}[!ht]
\centering
\renewcommand\arraystretch{1.3}
\begin{tabular}{c || c | c | c  |c }
& \multicolumn{4}{c}{Total levels of the grid hierarchy} \\
    & $2$ levels  & $3$ levels & $4$ levels, $9$ GL nodes & $4$ levels, $17$ GL nodes   \\ \hline\hline 
$\Delta t [ \times 10^{-7}$s$]$  & $1.4$ & $1.4$  & $0.7$ & $1.4$ \\\hline
CPU Time $[$s$]$ & $0.42$ & $0.78$ & $0.87$ & $1.18$ \\ 
\end{tabular}
\caption{Global time-step $\Delta t$ and  mean wall-clock CPU time by iteration for each simulation performed over different mesh and Gauss-Lobatto nodes hierarchy.}
\label{tab:1D_PMF_delta_T}
\end{table}

\subsection{Two-dimensional dimethyl ether jet}
\label{subsec:2D_DME_jet}

The simulations presented in the previous sections are simplified test cases to
assess the numerical accuracy and stability of the AMLSDC strategy, as well as to validate the physical behavior
for reacting flows.
Here we present a more complex case,
simulation of a reacting two-dimensional dimethyl ether (DME) jet.
DME is a surrogate for oxygenated fuels such as those produced from biomass.
DME is numerically challenging to simulate because the chemistry involved is extremely stiff and would require very small  time-steps for
purely explicit time-integration schemes, leading to unpractical requirements; however,
the AMLSDC strategy developed in the present paper is well-suited to treating this type of combustion simulation.

The numerical set-up is similar to the simulations reported in \cite{Emmett:2014}. The two-dimensional computational domain consists of a square of dimensions $-0.00114$m$<x<0.00114$m and $0<y<0.00228$m. A skeletal $39$ species mechanism is used to model the dimethyl ether chemistry \cite{BHAGATWALA:2015}. A premixed jet of DME and nitrogen
surrounded by a weak co-flow of air flows into a preheated domain at 1525$K$ filled with air at an initial pressure
of $40$atm. The inflow pressure is also set to 40atm with temperature, velocity and species mole fractions given by
\begin{align}
T_0 &= \eta T_{\rm jet} + \left(1-\eta \right) T_{\rm air}, \label{eqn:jet_init_T} \\
u_{0x} &= 0, \\
u_{0y} &= \eta u_{\rm jet} + \left(1-\eta \right) u_{\rm air}, \\
X_0 &= \eta X_{\rm jet} + \left(1-\eta \right) X_{\rm air}, \label{eqn:jet_init_X}
\end{align}
where $T_{\rm jet}=400$K, $T_{\rm air}=1525$K, $u_{\rm jet}=51.2$m.s$^{-1}$, and $u_{\rm air}=0.1 u_{\rm jet}$. The profile of the jet is controlled by the parameter $\eta$, which is given by
\begin{equation}
\eta = \frac{1}{2} \left(\tanh \frac{x+x_0}{\sigma} - \tanh \frac{x-x_0}{\sigma}  \right),
\end{equation}
where $x_0=5.69\times 10^{-5}$m and $\sigma=0.5 x_0$. The species mole fractions for the jet and air states are set to zero, except that
\begin{align}
X_{\rm jet}\left({\rm CH}_3 {\rm OCH}_3 \right) &= 0.2, \\
X_{\rm jet}\left({\rm N}_2 \right) &= 0.8, \\
X_{\rm air}\left({\rm O}_2 \right) &= 0.21, \\
X_{\rm air}\left({\rm N}_2 \right) &= 0.79.
\end{align}

An inflow boundary is used at the lower $y$-boundary, whereas characteristic outflow boundary conditions are applied at the other three boundaries. A perturbation is imposed to the jet through the application of a sinusoidal variation in the inflow velocity:
\begin{equation}
u_y \left(x,t\right) = u_{0y} + \tilde{u}\eta \sin \left(\frac{2 \pi}{L_x}x \right)\sin\left(\frac{2 \pi}{L_t}t\right),
\end{equation}
where $\tilde{u}=10$m.s$^{-1}$, $L_x = 0.00228$m and $L_t=10^{-5}$s.

\begin{figure}[!ht]
 \begin{subfigmatrix}{3}
  \subfigure[]{\includegraphics[height=0.75\textwidth]{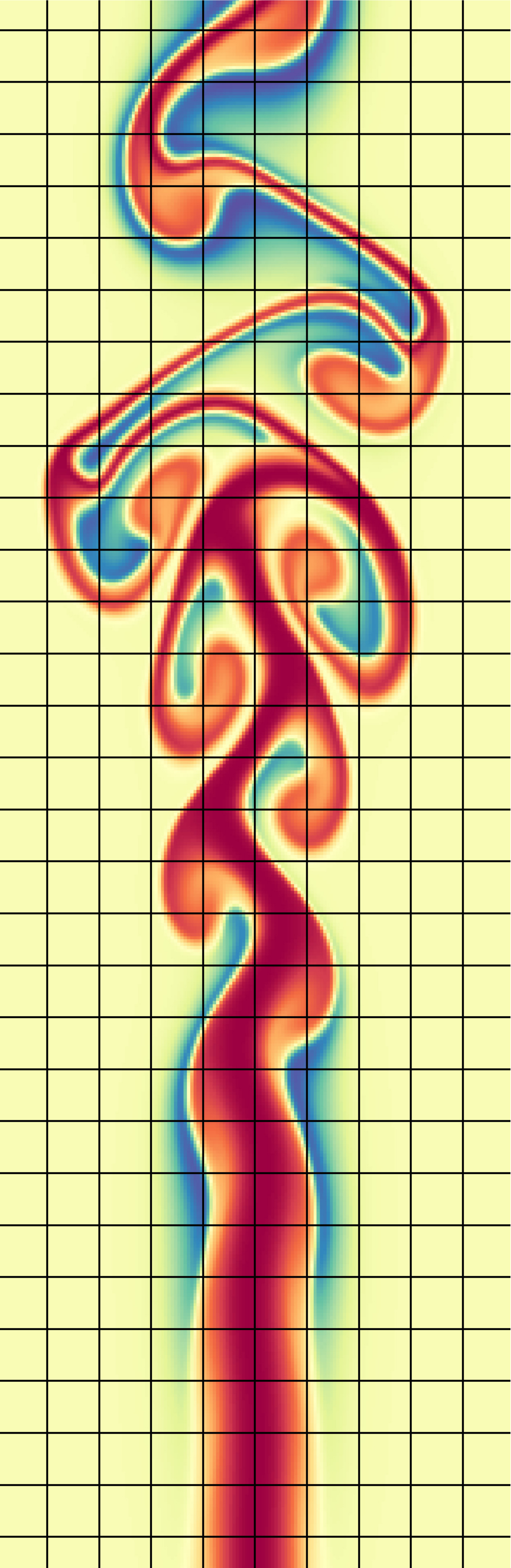}} 
  \hspace{-5cm}
  \subfigure[]{\includegraphics[height=0.75\textwidth]{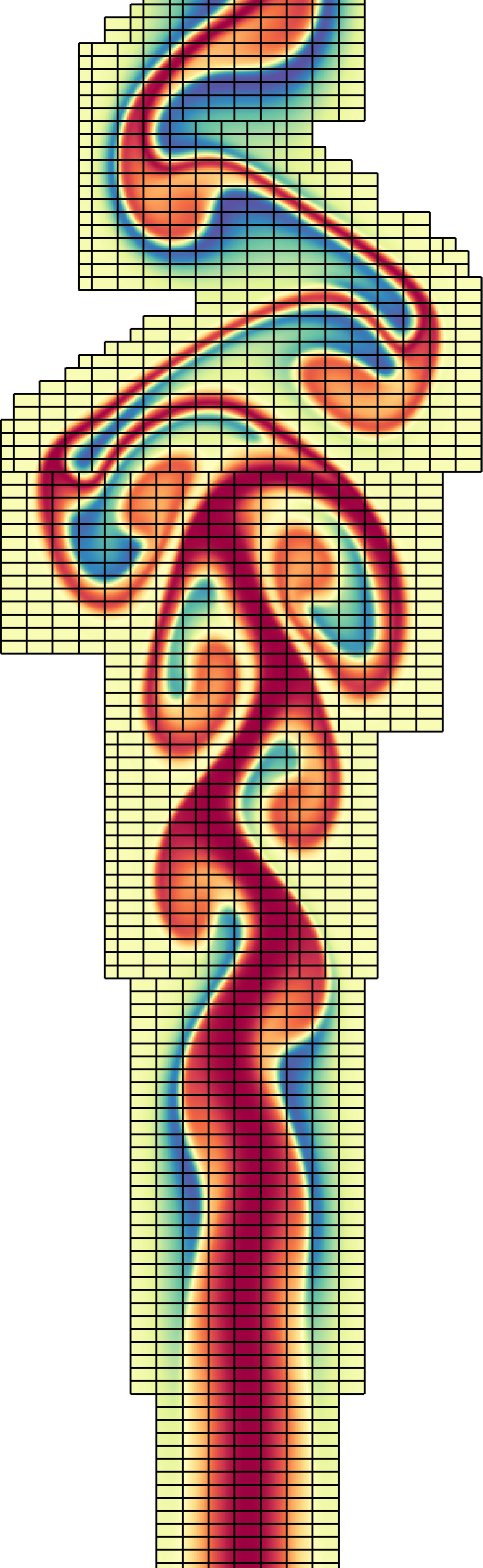}}
  \hspace{-5cm}
  \subfigure[]{\includegraphics[height=0.75\textwidth]{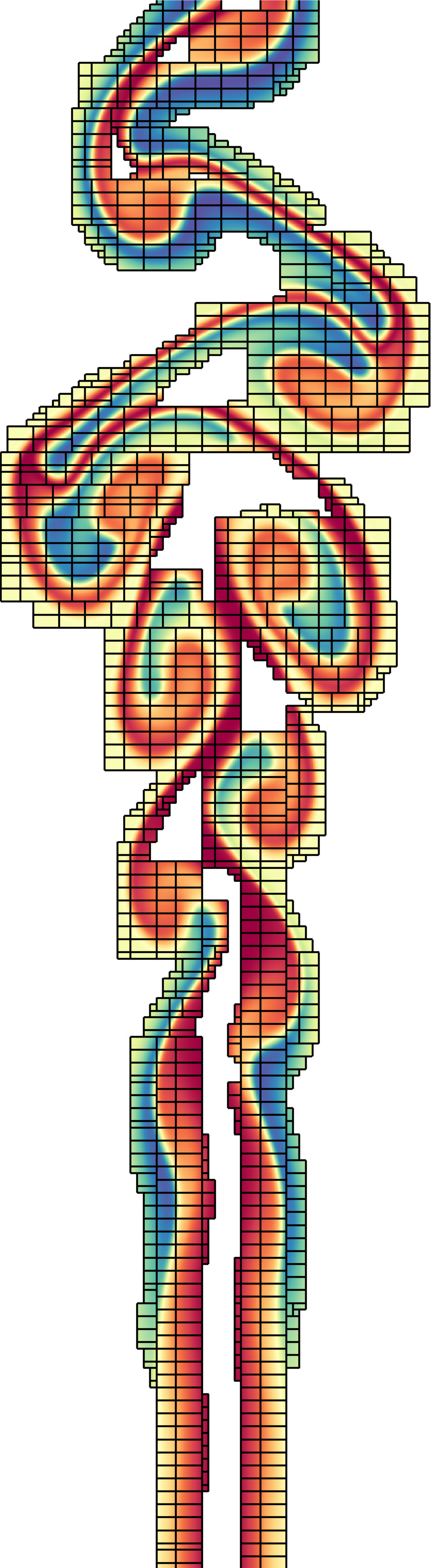}}
  \end{subfigmatrix}
 \caption{Instantaneous temperature field at $5 \times 10^{-5}$s, and corresponding grid hierarchy:  (a) baseline mesh, (b) first level of mesh refinement, (c) second level of mesh refinement.}
 \label{fig:2D_DME_Jet_AMR}
\end{figure}

The computational domain is discretized with a base mesh of $512 \times 512$ cells with a maximum of $2$ additional levels of mesh refinement.
Two criteria for adaptive mesh refinement are employed; namely, all cells where the magnitude of the vorticity $|\omega| > 5 \times 10^5$s$^{-1}$  or $Y_{\rm H} > 5 \times 10^{-6}$ will be tagged for refinement. The simulation is run for a physical time of $5 \times 10^{-5}$s. Figure~\ref{fig:2D_DME_Jet_AMR}
presents the instantaneous temperature field at $5 \times 10^{-5}$s, together with the corresponding grid hierarchy of $3$ levels of mesh. Note that for clarity, only the portion of the domain immediately surrounding the jet is shown.
The uniform baseline mesh is shown in Figure~\ref{fig:2D_DME_Jet_AMR}a. Here, each black box corresponds to a $16\times 16$ grid.
Figures~\ref{fig:2D_DME_Jet_AMR}a and \ref{fig:2D_DME_Jet_AMR}b
show the level 1 and level 2 grids, respectively, illustrating how dynamic refinement adapts to the solution.
Only the regions featuring significant vortical structures or chemical reactions are refined,
resulting in significant computational savings by using a coarse grid in parts of the domain where
higher resolution is not needed. 

The AMLSDC simulation with the {\tt RNS} code are compared to results of low-Mach-number code {\tt PeleLM} discussed
above \cite{Nonaka:2018}
and a high-order DNS code published in\cite{Emmett:2014}. 
For this low Mach number flow ($M \approx 0.13$) we expect acoustic waves to have a negligible effect on
the dynamics.
The initial set-up in the low-Mach-number {\tt PeleLM} code is the same as with {\tt RNS}, except that the domain is discretized on only one level with a mesh resolution of $ 2048 \times 2048$ points, which corresponds to the finest level of refinement in the {\tt RNS} simulation.

\begin{figure}[!ht]
 \begin{subfigmatrix}{4}
  \subfigure[$T$ {\tt PeleLM}]{\includegraphics[height=0.75\textwidth]{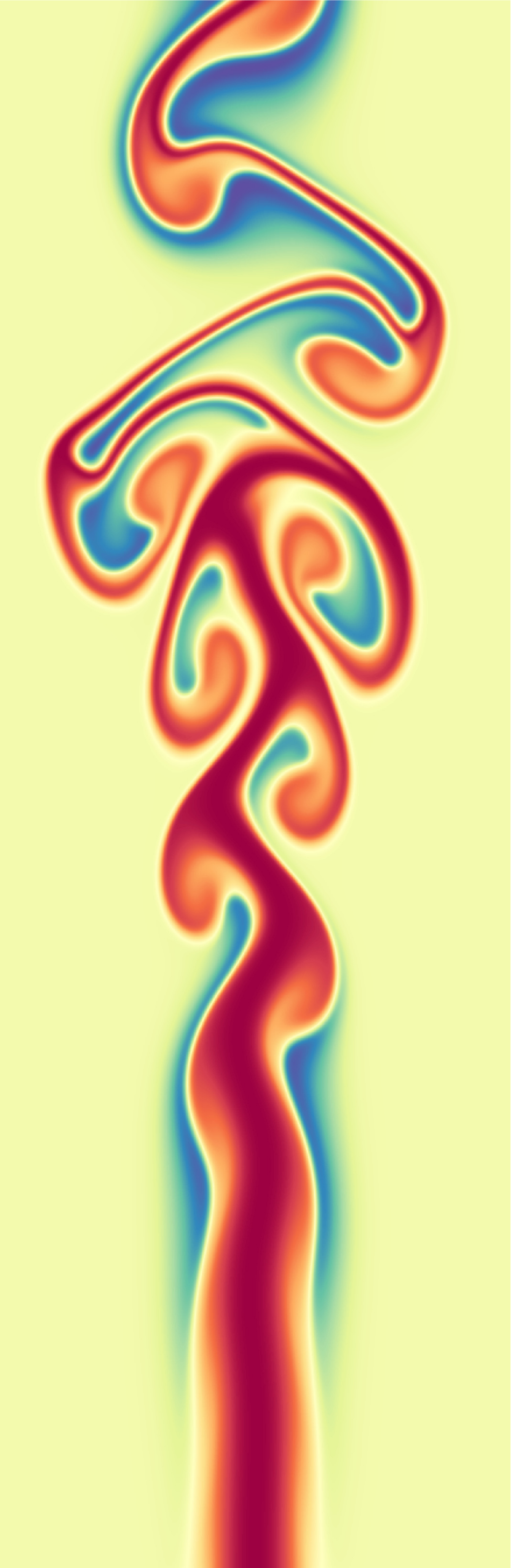}}
  \hspace{-0.9cm}
  \subfigure[$T$ {\tt RNS}]{\includegraphics[height=0.75\textwidth]{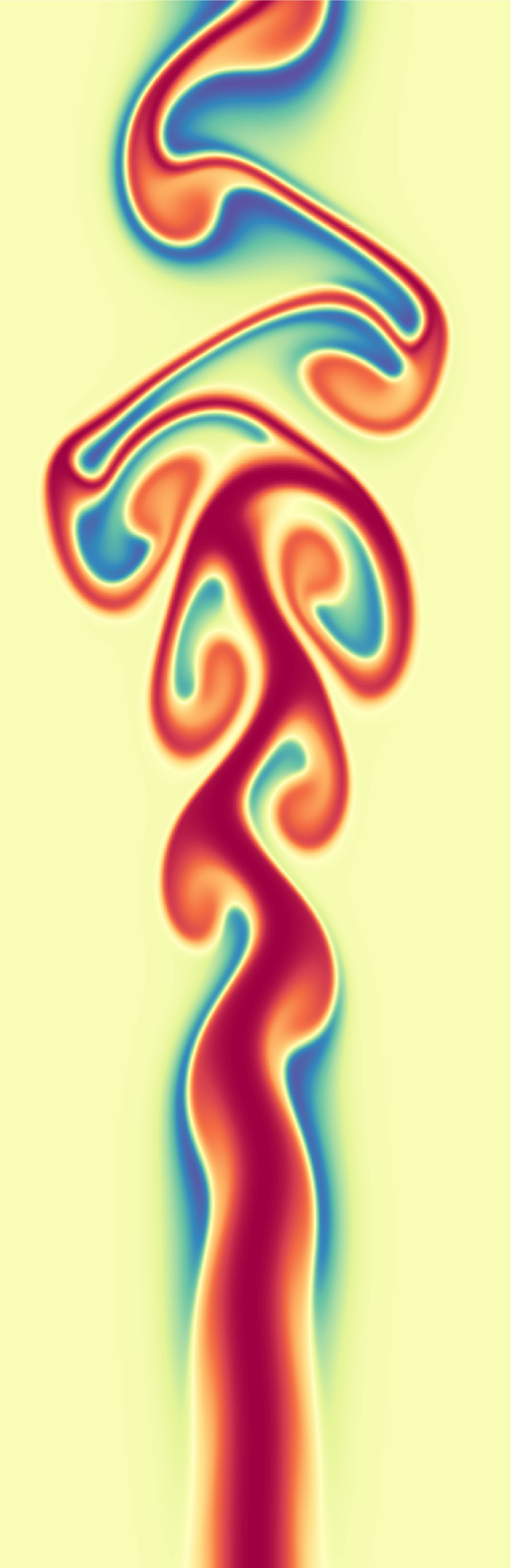}}
  \hspace{0cm}
  \subfigure[$Y_{\rm HCO}$ {\tt PeleLM}]{\includegraphics[height=0.75\textwidth]{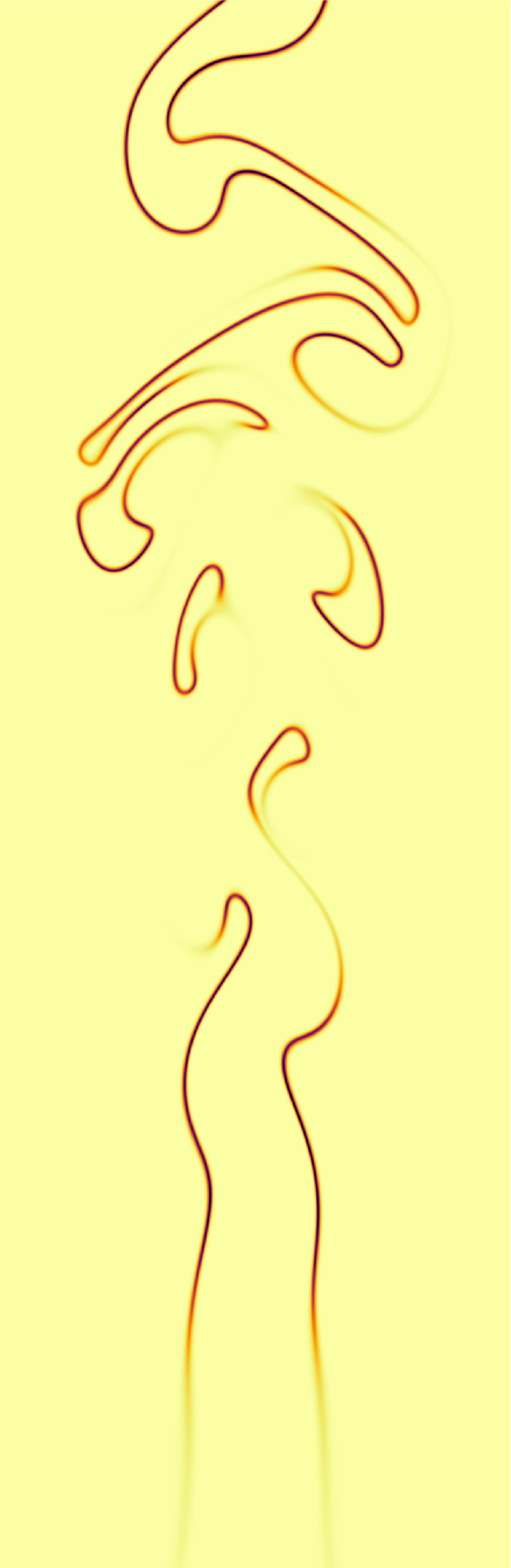}}
  \hspace{-0.9cm}
  \subfigure[$Y_{\rm HCO}$ {\tt RNS}]{\includegraphics[height=0.75\textwidth]{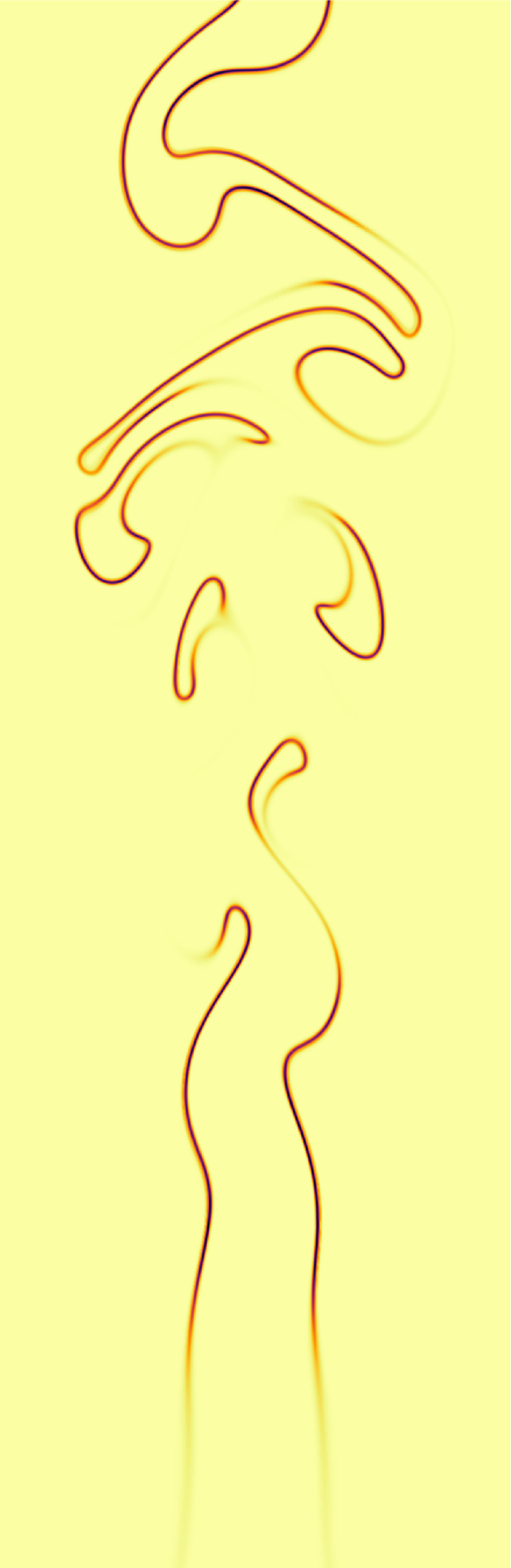}}
  \end{subfigmatrix}
 \caption{Instantaneous fields of temperature (panels (a) and (b)) and $Y_{\rm HCO}$ (panels (c) and (d)) at the physical time $t=5 \times 10^{-5}$s. Simulations performed with the low-Mach-number code {\tt PeleLM} are depicted in panels (a) and (c), while simulations with the {\tt RNS} code are in depicted panels (b) and (d).}
 \label{fig:2D_DME_Jet_RNS_PELELM}
\end{figure}

Qualitative comparisons are presented in Figure.~\ref{fig:2D_DME_Jet_RNS_PELELM}.
Instantaneous fields of temperature and $Y_{\rm HCO}$ at the physical time $t=5 \times 10^{-5}$s are shown on the left (a and b) and right (c and d) panels, respectively.
Simulations performed with the low-Mach-number code {\tt PeleLM} are shown in
Figure~\ref{fig:2D_DME_Jet_RNS_PELELM}a and \ref{fig:2D_DME_Jet_RNS_PELELM}c,
and simulations with {\tt RNS} code are in presented in 
Figure~\ref{fig:2D_DME_Jet_RNS_PELELM}b and \ref{fig:2D_DME_Jet_RNS_PELELM}d.
The results of the two codes are virtually indistinguishable in spite of the difference between the two physical formulations and numerical method that are used.
(High-order DNS results, not shown, show the same level of agreement.)
These results show that the AMLSDC strategy, coupled with an IMEX treatment for the evaluation of combustion, is able to treat a complex reacting flow with stiff chemistry.

\subsection{Three-dimensional hydrogen-air jet}
\label{subsec:3D_methane_jet}

Finally, we present a computationally intensive 3D AMR simulation.
The numerical set-up is similar to the 2D simulation in \S\ref{subsec:2D_DME_jet}. The computational domain is a cube of dimensions $-0.0015$m$<x<0.0015$m, $-0.0015m<y<0.0015$m and $0 <z<0.003$m. A $9$ species and $21$ reaction mechanism is used to model the hydrogen chemistry \cite{LiDryer:2004}.
Similar to the DME problem, a premixed jet of hydrogen and air
surrounded by a weak co-flow of air
flows into the domain filled with air at $p=$10atm and $T=$1300K. 
The inflow pressure is set to $10$atm, with temperature, mean inflow velocity and species mole fractions are set according to Eqs.~(\ref{eqn:jet_init_T})-(\ref{eqn:jet_init_X}), with $T_{\rm jet}=400$K, $T_{\rm air}=1300$K, $u_{\rm jet}=100$m.s$^{-1}$ and $u_{\rm air}=0.1 u_{\rm jet}$. Note that the initial mean velocity is set in the $z$ direction, the other components are set to zero. The profile of the jet is controlled by the parameter $\eta$, which is given by
\begin{align}
\eta = \frac{1}{4} \left(1-\tanh \frac{r-r_0}{\sigma_r} \right)\left(1-\tanh \frac{z-z_0}{\sigma_z} \right) \hspace{0.5cm} & \mathrm{if} z-z_0 < 5\sigma_z  , \\
\eta = 0 \hspace{0.5cm} & \mathrm{otherwise}.
\end{align}
Here, $r=\sqrt{x^2+y^2}$, $r_0=1.5\times 10^{-4}$m, $z_0=4\times 10^{5}$m, $\sigma_r=0.1 r_0$ and $\sigma_z=1.3 \times 10^{-5}$.
A time-varying turbulent field is superimposed to the inflow velocity $u \left(x,y,t\right)$ to perturb  the jet.
The species mole fractions for the jet and air states are set to zero, except that
\begin{align}
X_{\rm jet}\left({\rm H}_2 \right) &= 0.7, \\
X_{\rm jet}\left({\rm N}_2 \right) &= 0.3, \\
X_{\rm air}\left({\rm O}_2 \right) &= 0.21, \\
X_{\rm air}\left({\rm N}_2 \right) &= 0.79.
\end{align}

The computational domain is discretized with a base mesh of $128 \times 128 \times 128$ cells with a maximum of $2$ additional levels of mesh refinement. Three criteria for adaptive mesh refinement are employed; namely, all cells where the magnitude of the vorticity $|\omega| > 2 \times 10^7$s$^{-1}$,  or $Y_{\rm H} > 1 \times 10^{-5}$, or $\nabla T > 200$, will be tagged for refinement.
An inflow boundary is used at the lower $z$-boundary, whereas characteristic outflow boundary conditions are applied at the other three boundaries.

\begin{figure}[!ht]
 \begin{subfigmatrix}{2}
  \subfigure[]{\includegraphics[height=12cm]{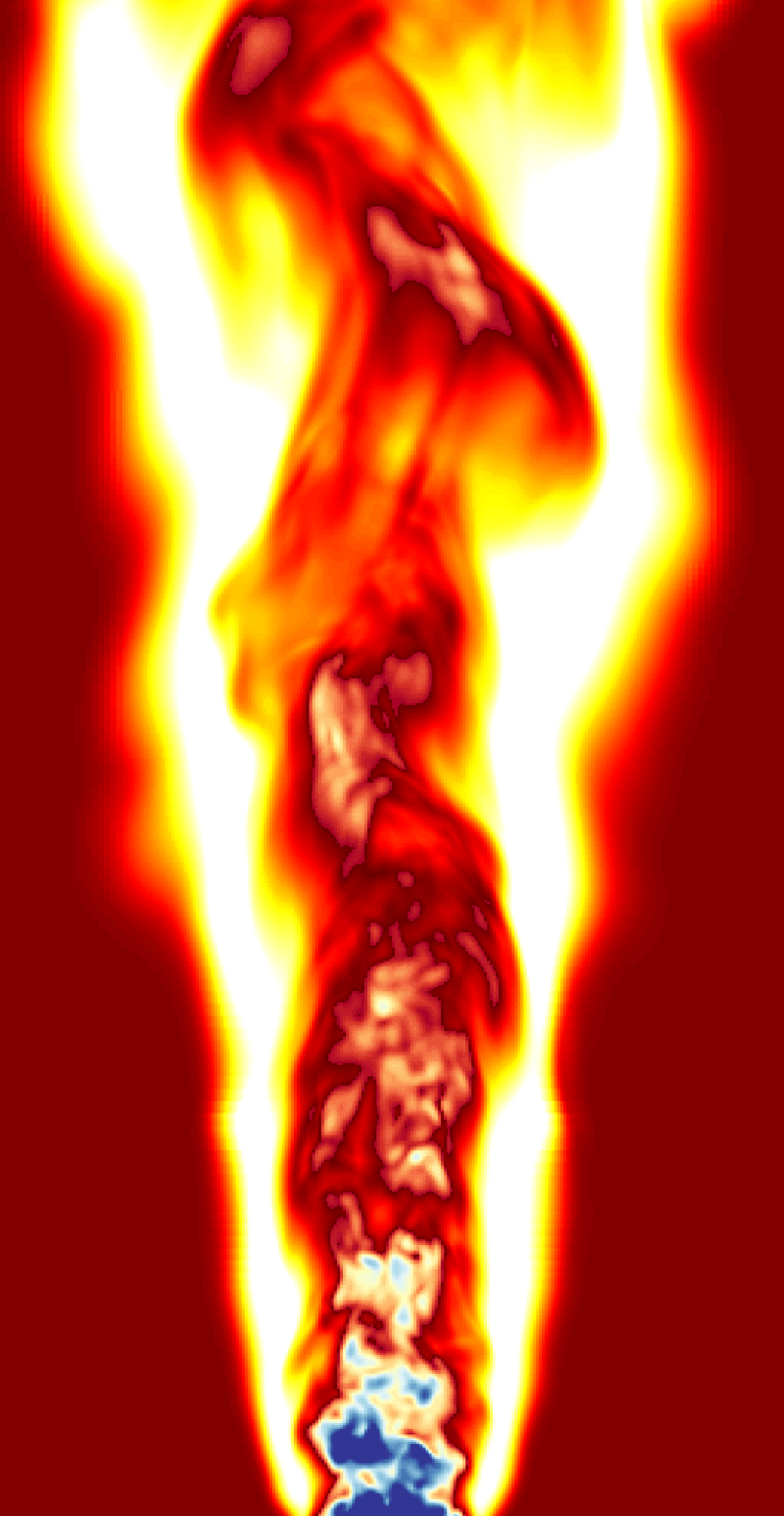}}
  \subfigure[]{\includegraphics[height=12cm]{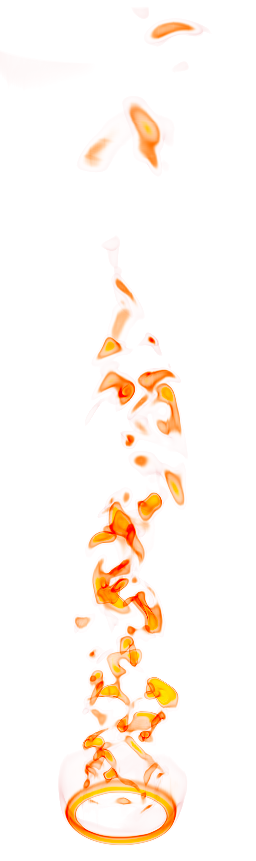}}
  \end{subfigmatrix}
 \caption{Three-dimensional simulation of a reacting hydrogen-air jet. Panel (a): planar projection of the 3D temperature field weighted by the magnitude of vorticity. Panel (b): volume rendering of $Y_{{\rm H}_2{\rm O}_2}$.}
 \label{fig:3D_methane_Jet_AMR}
\end{figure}

Figure~\ref{fig:3D_methane_Jet_AMR}
presents instantaneous results at nearly $9 \times 10^{-5}$s, where panel (a) is a planar projection of the 3D temperature field weighted by the magnitude of vorticity, and panel (b) is a volume rendering of $Y_{{\rm H}_2{\rm O}_2}$.

At the beginning of the simulation, the flame is not present inside the domain and no additional levels of mesh refinement are needed,
leading to a very low computational cost. Similar to the previous 2D simulation presented in \S\ref{subsec:2D_DME_jet}, the farther
the flame  propagates into the domain, the more of the grid is refined to resolve the flame with an appropriate level of discretization.
Once the flame is fully established, the finest level occupies about 11\% of the domain. 
Even at this relatively low percentage, approximately 87\% of the work is on the finest level.
With 512 cores, the time for one CPU to advance one zone for a time step is approximately 2.2 milliseconds on Cori 
at NERSC.
Strong scaling to 8192 cores when there are only roughly $12^3$ zones per core only increases this time to 3.2 milliseconds,
illustrating the good scalability of the algorithm.
These results show that the AMLSDC strategy, coupled with an IMEX treatment for the evaluation of combustion and WENO-based schemes for the spatial discretization, is able to be effectively utilize high performance computing to simulate a complex three-dimensional turbulent reacting flow.

\section{Conclusion}
\label{sec:conclusion}

We have presented a new fourth-order in space and time block-structured adaptive mesh refinement algorithm for the
reacting compressible Navier-Stokes equations.  The spatial discretization uses a higher-order finite volume treatment of
advection and diffusion.  The advective terms are treated with a fourth-order finite volume method using WENO reconstructions for forming flux values.  The method uses an implicit / explicit
SDC temporal integration strategy that treats advection and diffusion explicitly while treating reactions implicitly, enabling
the methodology to handle stiff reaction kinetics.

A key feature of the new methodology is the introduction of a new approach to time-stepping on an adaptive mesh hierarchy, referred
to as AMLSDC.  This new approach sweeps through all levels of the grid hierarchy in a fashion analogous to a multigrid V-cycle.
A FAS correction term is included in the discretization so that the coarse-solution accurately reflects the behavior on the fine 
grid. 
Using the solution on coarser levels to provide an approximation to the solution on finer levels reduces the computational
effort
compared to a traditional AMR time-stepping algorithm by reducing the number of SDC sweeps needed to achieve a given level of accuracy.
For the reacting flow cases presented here, the AMLSDC iteration typically converges in two V-cycles, resulting
in significantly fewer function evaluations on the fine grid than the traditional approach.
The more direct coupling between coarse and fine grids also leads to improved accuracy at coarse / fine grid boundaries and
avoids any order reduction.

Numerical examples of both non-reacting and reacting 
flow demonstrated the fourth-order convergence of the methodology in space and time.
We also validated the algorithm on a dimethyl ether jet flame, demonstrating that the 
new algorithm was able to match previously published numerical results.  Finally, we demonstrated the
utility of the new methodology for simulation of a turbulent hydrogen jet flame.
Overall, the algorithm developed here combines the utility of higher-order discretization and adaptive mesh
refinement, making it a potentially valuable approach for DNS of reacting flows.

\section*{Acknowledgment and Funding}

The work here was supported by the U.S. Department of Energy, Office of Science,
Office of Advanced Scientific Computing Research, Applied Mathematics program under contract number DE-AC02005CH11231. Part of the simulations were performed using resources of the National Energy Research Scientific Computing Center (NERSC), a DOE Office of Science User Facility supported by the Office of Science of the U.S. Department of Energy under Contract No. DE-AC02-05CH11231.

\bibliographystyle{tfq}
\bibliography{bib_manu_2017}

\end{document}